\documentclass[sigconf]{acmart}

\AtBeginDocument{%
  \providecommand\BibTeX{{%
    \normalfont B\kern-0.5em{\scshape i\kern-0.25em b}\kern-0.8em\TeX}}}

\copyrightyear{2022}
\acmYear{2022}
\setcopyright{acmcopyright}\acmConference[SIGMOD '22]{Proceedings of the 2022 International Conference on Management of Data}{June 12--17, 2022}{Philadelphia, PA, USA}
\acmBooktitle{Proceedings of the 2022 International Conference on Management of Data (SIGMOD '22), June 12--17, 2022, Philadelphia, PA, USA}
\acmPrice{15.00}
\acmDOI{10.1145/3514221.3526176}
\acmISBN{978-1-4503-9249-5/22/06}

\usepackage{graphicx}
\usepackage{balance}

\usepackage{xcolor}
\usepackage[most]{tcolorbox}

\usepackage{amsthm}
\usepackage{bbm}
\usepackage{subfigure}
\usepackage{url}
\usepackage{booktabs}
\usepackage{enumitem}
\usepackage{booktabs}
\usepackage{caption}
\usepackage{makecell}
\usepackage{multirow}
\usepackage{tcolorbox}
\usepackage{pifont}
\usepackage{xspace}
\newcommand{\sys}{\textsc{OnlineTune}\xspace}

\newfloat{figtab}{htb}{fgtb}
\makeatletter
  \newcommand\figcaption{\def\@captype{figure}\caption}
  \newcommand\tabcaption{\def\@captype{table}\caption}
\makeatother

\usepackage{enumitem}
\usepackage{pifont}
\usepackage[ruled, vlined, linesnumbered]{algorithm2e}
\usepackage{amsmath}
\usepackage{booktabs}
\usepackage{multirow}
\newcommand{\R}{\mathbb{R}}
\usepackage{adjustbox}
\usepackage{graphicx}
\usepackage{float}
\usepackage{xcolor}
\SetKwInOut{Parameter}{Parameters}
\usepackage{hyperref}
\def\Snospace~{Section {}}

\usepackage{xspace}

\begin{document}

\title{Towards Dynamic and Safe Configuration Tuning for Cloud Databases}

\author{Xinyi Zhang}
\authornote{School of CS \& Key Laboratory of High Confidence Software Technologies, Peking University}
\authornote{Database and Storage Laboratory, Damo Academy, Alibaba Group}
\authornote{Center for Data Science, Peking University \& National Engineering Laboratory for Big Data Analysis and Applications}

\affiliation{%
  \institution{Peking University \& Alibaba Group}
}
\email{zhang\_xinyi@pku.edu.cn}

\author{Hong Wu}
\authornotemark[2]

\affiliation{%
  \institution{Alibaba Group }
  \streetaddress{Wangjing Tower A, Ali Center, Chaoyang District} 
  \postcode{100102} 
}
\email{hong.wu@alibaba-inc.com}

\author{Yang Li}
\authornotemark[1]
\affiliation{%
  \institution{Peking University}
}
\email{liyang.cs@pku.edu.cn}

\author{Jian Tan}
\authornotemark[2]
\affiliation{%
  \institution{Alibaba Group}
  \streetaddress{Wangjing Tower A, Ali Center, Chaoyang District} 
  \postcode{100102} 
}
\email{j.tan@alibaba-inc.com}

\author{Feifei Li}
\authornotemark[2]
\affiliation{%
  \institution{Alibaba Group }
  \streetaddress{Wangjing Tower A, Ali Center, Chaoyang District} 
  \postcode{100102} 
  }
\email{lifeifei@alibaba-inc.com}

\author{Bin Cui}
\authornotemark[1]
\authornotemark[3]
\authornote{Institute of Computational Social Science, Peking University (Qingdao)}
\affiliation{%
  \institution{Peking University}
  \postcode{100080}}
\email{bin.cui@pku.edu.cn}

\renewcommand{\shortauthors}{Xinyi Zhang and Hong Wu, et al.}

\sloppy
 \fancyhead{} 



\begin{abstract} 
Configuration knobs of database systems are essential to achieve high throughput and low latency.
Recently, automatic tuning systems using machine learning methods (ML) have shown to find better configurations compared to experienced database administrators (DBAs).
However, there are still gaps to apply the existing systems in production environments, especially in the cloud. 
First, they conduct tuning for a given workload within a limited time window and ignore the dynamicity of workloads and data.
Second, they rely on a copied instance and do not consider the availability of the database when sampling configurations, making the tuning expensive, delayed, and unsafe. 
To fill these gaps, we propose \sys, which tunes the online databases safely in changing cloud environments.
To accommodate the dynamicity, \sys embeds the environmental factors as context feature and adopts contextual Bayesian Optimization with context space partition to optimize the database adaptively and scalably. 
To pursue safety during tuning, 
we leverage the black-box and the white-box knowledge to evaluate the safety of configurations and propose a safe exploration strategy via subspace adaptation.
We conduct evaluations on dynamic workloads from benchmarks and real-world workloads. Compared with the state-of-the-art methods, \sys achieves 14.4\%\textasciitilde 165.3\% improvement on cumulative performance while reducing 91.0\%\textasciitilde99.5\% unsafe configuration recommendations.

\end{abstract}



\maketitle

\begin{figure}
\centering
    \includegraphics{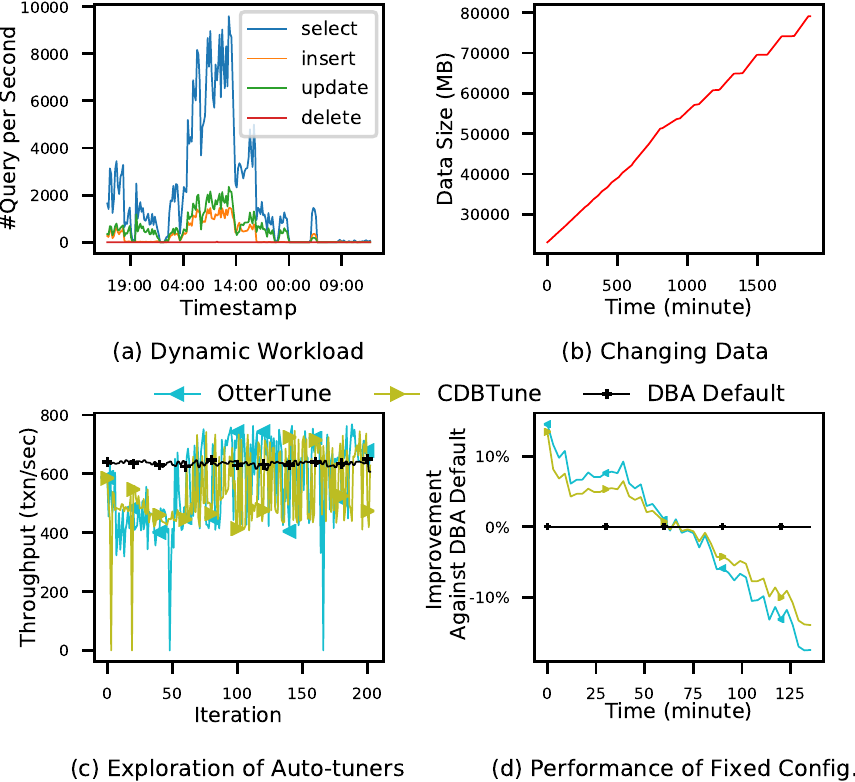}
\caption{Motivating Examples: Figures (a) and (b) show the dynamic environment in the cloud.
Figures (c) and (d) show the performance of existing auto-tuners.
In (c), the tuners tune a static workload with numerous unsafe trials.
In (d), we apply the best configurations found in (c) to a dynamic workload and observe decreasing improvement.}
\label{fig:motivation}
\end{figure}

\section{Introduction}

Modern database management systems (DBMS) have dozens of configurable knobs that control the runtime behaviors ~\cite{DBLP:journals/pvldb/DuanTB09} and impact database performance such as throughput and latency. 
Finding an optimal configuration for a given workload is proven to be an NP-hard problem ~\cite{DBLP:conf/sigmetrics/SullivanSP04} and it is impossible to enumerate and evaluate all the combinations. 
To find promising configurations, database administrators (DBAs) put considerable effort into tuning the configurations based on their experience.
In addition,  the workloads from real-world applications are {\em dynamic}, which means the properties of the workloads vary over time. 
Figure \ref{fig:motivation} (a) shows a dynamic workload trace from a real-world application. 
Similarly, the underlying data size and distribution constantly change due to the data modification operations. 
In Figure \ref{fig:motivation} (b), the size of underlying data increases almost three times after running TPC-C workload for 33 hours. 
The  {\em dynamicity} of the environment makes configuration tuning more difficult.

Many recent studies focus on automatic configuration tuning for DBMSs using Machine Learning (ML) techniques, including Gaussian Process Regression from OtterTune\cite{DBLP:conf/sigmod/AkenPGZ17}, and Deep Deterministic Policy Gradient from CDBTune~\cite{DBLP:conf/sigmod/ZhangLZLXCXWCLR19} and QTune~\cite{DBLP:journals/pvldb/LiZLG19}. 
These ML-based tuning systems can efficiently sample more promising configurations than DBAs; that is, they could find near-to-optimal configurations with fewer trials.
They utilize ML models to suggest promising configurations and update the models by evaluating the configurations iteratively.
To achieve this, they first need to copy the target DBMS instance to either replay or forward the workload. 
In each iteration, the tuner suggests a configuration and applies it to the copied DBMS.
Then, the workload is executed to evaluate the performance metrics of the suggested configuration, and the model is updated based on the evaluation. 
This process repeats until the models converge or stop conditions are met (e.g., budget limits).

Unfortunately, there are several major problems when deploying these systems in production, especially in the cloud. 
All the existing systems fall into {\em offline tuning} systems since the tuning process runs in a copied instance using a workload trace within a certain period. 
On the one hand, they only consider part of the workload, assuming the workload pattern never shifts too far.
The tuning process usually takes hours or days to find near-to-optimal configurations.
Consequently, the offline methods can not adapt to the dynamic workload from real-world applications. 
On the other hand, these systems rely on a copied instance and require the infrastructure to replay the workloads~\cite{DBLP:conf/sigmod/YanJJVL18}. 
For the consideration of data privacy, the copied instance should be launched in the users' environment (e.g., Virtual Private Cloud). 
Every time end-users want to tune the configurations, they should purchase a copied instance and prepare the workload and data based on a snapshot. 
This cloning process is time-consuming makes the tuning task expensive and delayed. 
The Total Cost of Ownership (TCO)  is increased for end-users, which is an important factor in the cloud.

To fill the above gaps, it is desirable to design a configuration tuning system coordinating with the online database to adapt to the workload changes directly and conveniently. 
Tuning configuration knobs for an online database does not need to copy the instance, so the TCO issue is directly solved. 
However, the online tuning system should address the following challenges. 
First, high availability is essential to online databases, and it is not allowed to sample bad configurations during tuning, causing the performance downgrade. 
However, it is inevitable for existing approaches to sample bad configurations.
Figure \ref{fig:motivation} (c) shows the process of existing automatic configuration tuning systems tuning the TPCC workload in a MySQL instance. OtterTune utilizes Bayesian Optimization to suggest promising configurations by balancing exploration and exploitation, and CDBTune adopts reinforcement learning to learn the tuning policy via trials-and-error. 
They could find near-to-optimal configurations. 
However, 50\%\textasciitilde70\% of their recommended configurations are worse than the default. 
The tuning even causes two system hangings by configuring the total memory (e.g., buffer pool, insert buffer, sort buffer, etc.) larger than the machine's physical capacity. 
Such harmful recommendations would expose enormous risks to online databases.
Second, the dynamic environment in the cloud should be considered, and we can not directly leverage the offline tuning approaches. 
Figure \ref{fig:motivation} (d) shows the performance of applying the best configurations obtained from the tuning processes in Figure \ref{fig:motivation} (c) to a dynamic workload (changing the transaction weight gradually from the original TPC-C workload). 
We observe that the applied configurations have better performance at the beginning. 
However, the configurations recommended by the offline tuning systems mismatch the dynamic workload afterward.
They become worse than the default after 75 minutes.
From these perspectives, the online database tuning should fulfill the following desiderata:

 \noindent\textbf{Dynamicity}: The tuner is capable of responding to the dynamic environment (e.g., workload and its underlying data) adaptively.
 
 \noindent\textbf{Safety}: The tuner should recommend configurations that do not downgrade the database performance during the tuning process.

In this paper, we propose \sys, an online tuning system that tunes the databases safely and adaptively in the constantly changing cloud environment. 
For \textbf{dynamicity}, we formulate the online tuning problem into a contextual bandit problem. 
\sys featurizes the context and optimizes the database performance over a context-configuration joint space.
We propose a clustering and model section strategy to scale up \sys with accumulated observations. 
\sys clusters the observations, fit multiple models, and selects the appreciate one for a given context. 
For \textbf{safety}, \sys safely explores the configuration space.
It evaluates the safety of configurations by leveraging both the black-box knowledge (i.e., posterior estimate from the model) and white-box knowledge (i.e., heuristics-based rules from the domain experience). 
Since satisfying the safety constraint in the continuous and high-dimension space is non-trivial~\cite{DBLP:conf/icml/KirschnerMHI019}, we transform the high-dimension optimization problem into a sequence of subspace problems that can be solved efficiently.
Each model maintains a configuration subspace centered around the best configuration found so far.
\sys  starts from configurations similar to those known to be safe and expands the subspace to facilitate further explorations. 
Specifically, we make the following contributions:

 \begin{itemize}[leftmargin=*]
 \item To address the challenges in real DBMS scenarios with the dynamic workload, we define the online tuning problem and solve it as a contextual bandit problem with safety constraints. 
 To the best of our knowledge, \sys is the first online configuration tuning system for DBMS with safety consideration.
 \vspace{-1em}
 \item We propose a context featurization model that extracts features of workloads and underlying data. 
 Using this, \sys adopts the contextual Bayesian optimization technique to optimize the database adaptively with constantly changing environments.
 
 \item To enhance the scalability of \sys with extensive data in the cloud, we propose a clustering and model selection strategy that significantly decreases the computation complexity.

 \item To solve the safety issue, we combine the black-box and the white-box knowledge to evaluate the safety of configurations and propose a safe exploration strategy via subspace adaptation, largely reducing the risks of applying harmful configurations. 

 \item We implement the proposed method and evaluate on dynamic workloads from benchmarks and the real-world application. Compared with the state-of-the-art techniques, \sys achieves 14.4\%\textasciitilde 165.3\% improvement on cumulative improvement while decreasing 91.0\%\textasciitilde99.5\% unsafe recommendations.
 \end{itemize}

The remainder of the paper is organized as follows. 
We review the related works in ~\autoref{sec:related} and formally define the online tuning problem in ~\autoref{sec:problem}. 
A system overview of \sys is presented in ~\autoref{sec:sys}, followed by a description of our techniques for contextual performance modeling in ~\autoref{sec:context}, safe configuration recommendation in ~\autoref{sec:safe}. ~\autoref{sec:exp} presents our experimental evaluation.
Finally, we conclude in ~\autoref{sec:conclusion}. 
\section{Related Work}\label{sec:related}


 \noindent\textbf{Configuration Tuning.} There has been an active area of research on tuning configurations of DBMS, which can be summarized as:
 \begin{itemize}[leftmargin=*]
\item\textbf{Rule-based.} 
Rule-based methods recommend configurations using heuristic rules. 
Database vendors develop tuning tools to provide DBA with knobs recommendations through identifying database's bottlenecks due to misconfigurations ~\cite{DBLP:conf/cidr/DiasRSVW05} or asking the DBA questions about their application ~\cite{DBLP:conf/vldb/WeikumMHZ02}. 
Wei et al. propose to generate fuzzy rules for database tuning~\cite{DBLP:conf/fskd/WeiDH14}. 
BestConfig~\cite{DBLP:conf/cloud/ZhuLGBMLSY17} searches  configurations based on several heuristics. 
Rule-based methods strongly depend on the assumptions of their heuristics and fail to utilize knowledge gained from previous tuning efforts.

\item\textbf{Learning-based.} 
iTuned ~\cite{DBLP:journals/pvldb/DuanTB09}, Ottertune~\cite{DBLP:conf/sigmod/AkenPGZ17} and ResTune~\cite{DBLP:conf/sigmod/ZhangWCJT0Z021} use  Bayesian Optimization (BO) based method, modeling the tuning as a black-box optimization problem. 
Reinforcement Learning (RL) is adopted in ~\cite{DBLP:conf/sigmod/ZhangLZLXCXWCLR19, DBLP:journals/pvldb/LiZLG19} to tune DBMS by learning a neural network between the internal metrics and the configurations. 
In the field of data analytic systems, ReIM~\cite{DBLP:conf/sigmod/KunjirB20} studies the problem of tuning the memory allocation and develops an empirically-driven algorithm. 
And Tuneful ~\cite{DBLP:conf/kdd/FekryCPRH20, DBLP:journals/corr/abs-2001-08002} combines incremental Sensitivity Analysis and BO to prone configuration space. 
Although RL methods can adapt to workloads by fine-tuning its neural network, additional time and evaluation samples are needed. 
All the above methods train a machine learning model to learn the offline tuning policy and cannot promptly respond to the dynamic environment. 
Besides, they do not consider the safety constraints when interacting with the database, which restricts them from being deployed in production.
\end{itemize}

\noindent
\textbf{Query Featurizing}.
Query featurizing aims at translating plain SQLs into their vectorized representations.
We summarize it into SQL text parsing and logical plan parsing based on the input.
SQL text parsing directly processes the query texts.
TF-IDF is used to represent a query as a collection of the weighted frequencies of its individual word tokens~\cite{DBLP:conf/simbig/MakiyamaRS15,DBLP:conf/sigmod/ZhangWCJT0Z021}.
However, its unbounded vocabulary makes generalization across workloads difficult.
Therefore ResTune restricts its calculation to reserved SQL keywords~\cite{DBLP:conf/sigmod/ZhangWCJT0Z021}.
To support larger vocabulary, ~\cite{DBLP:conf/cidr/JainYCH19,DBLP:conf/sigmod/ZolaktafMP20} resort to representation learning which is frequently used in NLP.
Representation learning produces dense vectors capturing nuanced relationships of unstructured data.
They use deep models, e.g., LSTM or CNN, to learn the distributional embeddings.
Logical plan parsing parses a query plan to aggregate key features, including the cost or categories of operators, scanned tables, and predicates.
It has been adopted by many state-of-the-art works in the field of query-performance prediction or cardinality estimation~\cite{DBLP:conf/cidr/AkdereCRUZ11, DBLP:conf/icde/AkdereCRUZ12, DBLP:conf/icde/GanapathiKDWFJP09, DBLP:conf/cidr/KipfKRLBK19, DBLP:journals/pvldb/MarcusNMZAKPT19,DBLP:journals/corr/abs-1905-06425,DBLP:journals/pvldb/SunL19,DBLP:journals/dase/LanBP21}.
QTune encodes query type, involved tables, query operators and the corresponding costs to predict the internal metrics of a database~\cite{DBLP:journals/pvldb/LiZLG19}.

\vspace{0.5em}
\noindent\textbf{Bayesian Optimization}. Our algorithm falls under the general umbrella of Bayesian optimization (BO). 
It learns and optimizes a black-box function over configuration space~\cite{DBLP:conf/nips/SnoekLA12}. 
BO works iteratively: (1) updating the  surrogate model that describes the relationship between 
configurations and their performances and (2) choosing the next configuration to
evaluate by computing the acquisition function value.
The acquisition function measures the utility of candidate points for the next evaluation by trading off the exploration of uncertain areas and exploiting promising regions.

BO has been extensively used in many scenarios, including  hyper-parameter tuning ~\cite{DBLP:conf/nips/BergstraBBK11, DBLP:conf/icdm/WistubaSS15, DBLP:conf/sigmod/LibertyKXRCNDSA20,DBLP:journals/isci/DuGSW22,DBLP:journals/chinaf/KaediGA13}, experimental design~\cite{DBLP:conf/nips/FosterJBHTRG19} and controller tuning ~\cite{DBLP:conf/icra/CalandraSPD14, DBLP:conf/icra/MarcoBHS0ST17, DBLP:conf/cluster/FischerGB15,DBLP:conf/ijcai/FiduciosoCSG019}.
Contextual BO considers the environmental conditions by augmenting the GP kernel with extra context variables and uses $CGP-UCB$ to select promising action ~\cite{DBLP:conf/nips/KrauseO11}. 
DBA Bandit~\cite{DBLP:conf/icde/PereraORB21} chooses a  set of indices from finite and discrete configuration space based on the context of indexed columns and derived statistics from database optimizer.
It achieves an $\tilde{O}(\sqrt{n})$ regret bound after playing $n$ rounds as a safety guarantee, implying that the per-step average cumulative regret approaches zero after sufficiently many steps. 
However, guaranteeing the safety of every step is still challenging but vital for mission-critical applications. 
Recently, 
Constrained BO is proposed to optimize a black-box function with unknown constraint~\cite{DBLP:conf/uai/GelbartSA14, DBLP:conf/icml/GardnerKZWC14, DBLP:journals/corr/LethamKOB17}. But the constraint is not considered safety-critical, and the algorithm is allowed to evaluate unsafe parameters. 
The main instance of safe optimization is SAFEOPT algorithm~\cite{DBLP:conf/icml/SuiGBK15}. 
However, its formulation relies on a discretization of configuration space, which hinders high-dimensional applications. 
Meanwhile, scaling up BO with high dimensions is another active area. 
Recent works propose local modeling~\cite{DBLP:conf/aistats/WangGKJ18,DBLP:conf/aistats/WangGKJ18,DBLP:conf/nips/ErikssonPGTP19,DBLP:conf/icml/KirschnerMHI019} and space partitioning ~\cite{DBLP:conf/aistats/WangSJF14, DBLP:conf/nips/Munos11}, which achieve strong empirical results in high dimensional problems. 
Motivated by the advances, \sys refines a configuration \textit{subspace}  that can be discretized efficiently.

\section{Problem Statement}\label{sec:problem}

 \begin{figure*}[t]
\centering
\scalebox{0.7}{
    \includegraphics{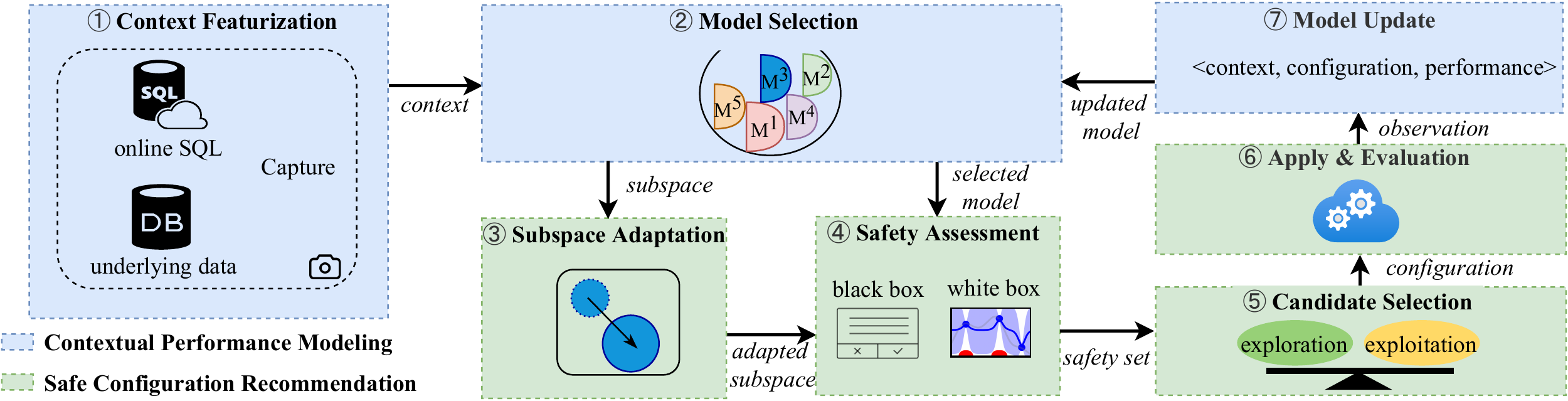}}
    \vspace*{-0.3cm}
\caption{\sys Workflow.}\label{fig:process}
\label{fig:dynamic-weight}
\vspace*{-0.3cm}
\end{figure*}

Consider a database system with a continuous configuration space $\Theta=\Theta_1\times  \Theta_2 \times ... \times \Theta_m$ and a context space $C$. 
The context $c \in C$ is uncontrollable environmental conditions, e.g., dynamic workloads.
We denote the database performance metrics as $f$, which can be any chosen metric to be optimized, such as {\em throughput}, {\em 99\%th percentile latency}, etc.
Given a configuration $\theta$ and context $c$, the corresponding performance $f(\theta, c)$ can be observed only after evaluation.


As we discussed, the online tuner should satisfy two requirements: \textbf{dynamicity} and \textbf{safety}. 
The \textbf{dynamicity} requirement means the tuner should consider changing environmental conditions (context) when recommending the configurations. 
We formalize it as a contextual bandit problem ~\cite{DBLP:conf/nips/KrauseO11}, where at each iteration $t$, the tuner receives context $c_t$ and outputs a configuration $\theta_t$ to maximize the payoff (i.e., database performance $f$).
The \textbf{safety} requirement indicates that we additionally need to ensure that, for each tuning iteration $t$, $f_t \geq \tau$ holds, where $f_t=f(\theta_t, c_t)$ and $\tau \in \R$ is a specific safety threshold.
We define the configurations that satisfy the above condition to be \textit{safe}.
The safety threshold indicates the degree of risk end-users can tolerate when adopting the tuning approach.
As the database is a mission-critical system, the service level agreement must be guaranteed by the cloud providers under the vendor default configuration ~\cite{DBLP:conf/sigmod/TaftESLASMA18, DBLP:journals/pvldb/TanZLCZZQSCZ19}. 
Intuitively, the safety threshold is set to the database performance under the default configuration (denoted as default performance). 
In case the default performance fluctuates with workload changes, we assume that the default performance for any given workload can be acquired.
Note in practice, the default performance  can be easily obtained or predicted when a historical knowledge base is available (such as in \cite{DBLP:conf/sigmod/AkenPGZ17,DBLP:conf/kdd/FekryCPRH20,DBLP:conf/sigmod/ZhangWCJT0Z021}).
For example, the user could train a regression model that inputs the context and outputs the default performance.  
Even without any previous knowledge, we can take some time to observe the default performance. 
At the very beginning, without any assumptions about $f$, searching the safe configurations is a nonsensical task. 
Hence, we also assume that, before the optimization, we are given an initial safety set that contains at least one safe configuration.  
Assuming that the objective is a maximization problem, the online tuning problem to solve for  $c_t \in C$ is :

\begin{equation}
\begin{aligned} \label{2}
& \mathop{\arg\max}_{\theta_t \in \Theta} f(\theta_t, c_t), \\
& \mbox{subject to}\
f(\theta_t, c_t) \geq \tau. \\
\end{aligned}
\end{equation}

\section{Overview of \sys}\label{sec:sys}
\noindent\textbf{Workflow.} 
To conduct online tuning, \sys first queries the default configuration and its performance to build an initial safety set.
The safety threshold is set to the default performance, as discussed in \autoref{sec:problem}.
Then, \sys functions iteratively.
It adapts to the dynamic environment by featurizing the context, recommending a promising configuration $\theta_t$ at the beginning of an iteration, and evaluating its performance during the iteration to update the model.
The interval size (i.e., time for one iteration) controls the granularity of \sys's adaptation to the dynamicity. 
Given a tuning time, \sys can collect more observations and make more fine-grained suggestions with a small interval size.
We use a three-minute interval by default and conduct sensitivity analysis in our experiments.
Within one iteration, the workflow of \sys forms an iterative cycle, as presented by \ding{172}-\ding{178} in Figure \ref{fig:process}.
The workflow consists of two stages: contextual performance modeling and safe configuration recommendation.  

\begin{figure*}
\centering
\includegraphics{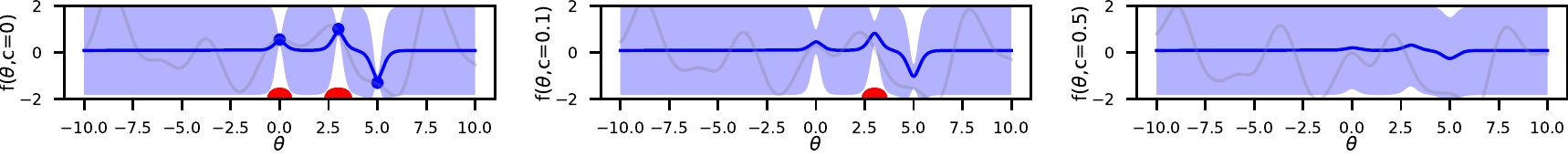}
\caption{Generalization over Contexts: \sys has the three observations (blue points) under the context $c=0$ (left) and fits a posterior distribution with mean (blue line) and confidence bound (light blue area). 
By exploiting correlations between different contexts, \sys can transfer knowledge of the objective function to a different context $c=0.1$ (middle), but little knowledge to the distant context (right). The red label shows the estimated safety set with zero as the safety threshold.}
\label{fig:gen-conext}
\end{figure*}
\underline{Contextual performance modeling} aims at obtaining a surrogate model that predicts the performance of given configurations in dynamic environments. 
The surrogate is a contextual Gaussian Process (GP) model fitted on historical observations (or the initial safety set, if no historical observations exist).
\ding{172} \sys first captures the dynamic factors (e.g., workload and its underlying data) through context featurization, obtaining a context (see \autoref{sec:context-feature} for details). 
\ding{173}
It then selects a contextual GP model fitted on the observations from the cluster with similar contexts.
The clusters are  periodically re-clustered in an offline manner (\autoref{sec:context-cluster}). \ding{178} The construction of contextual GP is introduced in \autoref{sec:context-model}.

\underline{Safe configuration recommendation} aims at selecting a safe and promising configuration from the configuration space. 
To avoid the aggressive exploration of BO, especially the over-exploration of boundaries, \sys
reduces the optimization over the whole configuration space into a sequence of \textit{subspace} optimization.
All operations in this stage are restricted in the \textit{subspace} where the safe optimization problem can be discretized and solved efficiently.  
In essence, the \textit{subspace} is centered around the best configuration estimated so far, gradually moving towards the optimal.
This stage inputs the selected surrogate model and its corresponding subspace.
\ding{174}
For a newly fitted model,  \sys initializes a subspace centered around the best-estimated configuration. 
Otherwise, \sys adapts the subspace according to the tuning history, e.g., expands the subspace when making consecutive successes (\autoref{sec:safe-update}). 
\ding{175} The adapted subspace is discretized to build a candidate set. 
\sys assesses the safety of the candidates based on the model's lower bound estimate, forming a safety set. 
It also consults the white box to dismiss unsafe configurations.
In case the heuristic white box excludes the optimal configurations from the safety set, \sys relaxes inappropriate rules (\autoref{sec:white}).
\ding{176} \sys then selects a configuration from the safety set by either maximizing the acquisition function or exploring the safe boundaries of subspace.
\ding{177} Finally, \sys applies the configuration to the online database and evaluates its performance.

\noindent\textbf{Architecture.}
The main parts of \sys are deployed in a backend tuning cluster (\sys sever), while the context featurization module is deployed in the database instance for data privacy concerns.
\sys server maintains a data repository that stores the historical observations from the previous tuning iterations, which can be initially empty.
\sys server interacts with the database instance via a controller that monitors the tuning tasks' states and transfers the data.

\section{Contextual Performance Modeling}
\label{sec:context}

\sys responds to dynamic environments when tuning online DBMSs.
It augments the GP kernel with context variables to learn the relationship between $\langle$context, configuration$\rangle$, and database performance.
We first discuss how to featurize context and construct a contextual GP model. 
Then, we present a clustering and model selection strategy to enhance the model's scalability.

\subsection{Context Featurization}\label{sec:context-feature}
Context featurization aims to capture uncontrollable dynamic factors, which affects the relationship between configurations and  database performances.
When tuning the database, the workload and underlying data are constantly and continuously changing due to the upstream applications and the DML statements (e.g., insert, delete and update).
\sys featurizes the two factors to adapt to the dynamicity.


\subsubsection{Workload Featurization}\label{sec:context-workload}
We now illustrate how to featurize the changing workload.
There are two dynamic aspects of workloads~\citep{DBLP:conf/sigmod/MaAHMPG18}: (1)  \textit{query arrival rate}:  the number of arriving queries per second can fluctuate. (2) \textit{query composition}: the types of queries may change, and the ratio of queries composition may vary.   

\textit{Query arrival rate} can be encoded by one dimension. For \textit{query composition}, we need to translate the plain queries into vectorized representations. 
 We adopt representation learning techniques~\cite{DBLP:journals/pami/BengioCV13} to extract informative encoding of queries and generalize across workloads.
 We choose LSTM, which have been used successfully for SQL query analysis ~\cite{DBLP:conf/sigmod/ZolaktafMP20,DBLP:conf/cidr/JainYCH19, DBLP:journals/corr/abs-1801-05613}. 
We use a standard LSTM encoder-decoder network ~\cite{DBLP:conf/nips/SutskeverVL14}  to ease the burden of collecting labeling data.
 The final hidden state on the encoder network provides a dense encoding for the query. 
 Lastly,  we average the query encoding, obtaining the \textit{queries composition} feature of a workload.

\subsubsection{Underlying Data Featurization}\label{sec:context-data}

Learning the  distribution of database data is a non-trivial task ~\cite{DBLP:journals/pvldb/YangLKWDCAHKS19, DBLP:journals/pvldb/HilprechtSKMKB20, DBLP:journals/pvldb/YangKLLDCS20}. 
Based on the observation that only the data changes affecting the workload queries are related to the tuning policy,
we use the following features from the DBMS optimizer: (1) estimate of rows to be examined by queries, (2) the percentage of rows filtered by table conditions in queries, (3) whether an index is used. 
The first two features are queries' cardinality estimation, capturing the effective changes in data size and distribution. 
The last feature indicates the index building/dropping operation.  
We average the three features of queries, obtaining the underlying data feature of a workload.

Finally, we concatenate the workload feature and underlying data feature to obtain the final context features. 
 Although query plans can provide extra information about the query execution, e.g.,  operators and costs, we do not encode them as contexts.
This is because the extra information is affected by \sys's previous configuration, which is unsuitable as a context for future tuning.
In addition, modern DBMSs have hundreds of plan operators~\cite{DBLP:conf/sigmod/0006ZJWBLMP21}, encoding the plan operators could cause the ``dimensionality issue'' when augmenting the sparse and high-dimensional variables to GP kernel.
Other dynamic factors could affect database performance, such as hardware configuration, data partition change, user-invoke configurations, etc.
Their changes usually occur intermittently.
\sys can re-initialize a tuning task when the changes occur or encode these factors.
For example, \sys can encode hardware configuration (e.g., memory size, \#CPU) to support hardware updates.
Database anomalies (e.g., out of disk space and cybercriminal attack) are out of \sys's scope.

\subsection{Performance Modeling with Contexts}\label{sec:context-model}
A performance model estimates the performance metrics given context and potential configurations. 
We adopt the contextual GP as the performance model, which extends the GP to support dynamic environments.
GP is a popular surrogate model for objective modeling in Bayesian Optimization~\cite{automl} due to its expressiveness and well-calibrated uncertainty estimates. 
GP provides confidence bounds on its predictions to model the objective function with limited samples. 
In such a scenario, other data-intensive techniques, e.g., deep learning~\cite{DBLP:conf/sigmod/ZhangLZLXCXWCLR19, DBLP:journals/pvldb/LiZLG19} may struggle with low data efficiency and interpretability~\cite{DBLP:conf/kdd/FekryCPRH20}. 
In this section, we focus on learning the performance model and leave the phase of safe tuning in \autoref{sec:safe}.

We aim to learn a performance model for different but correlated contexts.
In the learning phase, we construct the probability distribution $p\left(f|\theta, c, H_{t}\right)$ of the target function $f = f\left(\theta, c\right)$ given an unspecified input $\theta$, the observed context feature $c$ and the observations $H_t = \{c_i,\theta_i,y_i\}^t_{i=1}$. 
The posterior distribution over $f$ is a contextual GP with mean $\mu_t\left(\theta, c\right)$ and variance $\sigma^2_t\left(\theta,c\right)$:
\vspace{-0.2em}
\begin{equation}
\begin{aligned}
    \mu_t(\theta,c) &= k^T \left(K +\sigma^{2}I\right)^{-1} y_{1:t}, \\
    \sigma_t^2(\theta) &= k\big(\left(\theta, c\right), \left(\theta,c\right)\big) - k^T {\left(K+\sigma^{2}I\right)}^{-1} k,
\end{aligned}\label{equ:gp}
\end{equation}
where $k=[k((\theta_1,c_1),(\theta,c)),...,k((\theta_t,c_t),(\theta,c))]^T$ and K is a covariance matrix whose
$(i, j)$th entry is $K_{i,j}=k((\theta_i,c_i),(\theta_j,c_j))$. 
The kernel $k((\theta, c), (\theta^{'},c^{'}))$ should model the distances between points of configuration and context.
Concretely, we construct an additive kernel $k_{\Theta}(\theta, \theta^{'}) + k_C(c, c^{'})$. 
We use a linear kernel $k_C(c, c^{'})$ to model the dependence on contexts and a Martin kernel $k_\Theta(\theta, \theta^{'})$ to model the nonlinear performance on configurations.  
Intuitively, such design could model overall trends according to the context, and the configuration-specific deviation from this trend ~\cite{DBLP:conf/nips/KrauseO11}.

The composite kernel implies that function values are correlated when configurations and contexts are similar~\cite{DBLP:journals/corr/BerkenkampKS16}. We expect the same configuration across correlated contexts to have similar performance predictions. The correlations between contexts can significantly speed up the tuning. Figure \ref{fig:gen-conext} shows a simple scenario with a one-dimension context feature and a one-dimension configuration. 
Even though the algorithm has only explored the configuration space at the first context (z = 0, left figure), the correlation between the functions generalizes information to the unobserved context (z = 0.1, central figure). 
The knowledge transfer improves data efficiency and reduces the number of required evaluations.

\begin{algorithm}[t]
\DontPrintSemicolon
\KwIn{$H_t = \{c_i,\theta_i,y_i\}^t_{i=1}$}
\KwOut{Multiple contextual GPs and a decision boundary}
\caption{Offline Clustering\label{alg:context-space}}
\If{Need\_ReLearn()\label{alg:split}}
{Conduct DBSCAN on $\{c_i\}_1^{t}$ and get cluster labels $\{l_i\}_1^{t}$.\;\label{alg:cluster}
Fit Contextual GP  $p\left(f|c,\theta, H_{t}\right)$ for each clustering.\;\label{alg:context-fit}
Fit SVM model on $\{c_i,l_i\}_1^{t}$, obtaining decision boundary.\;\label{alg:context-svm}
}

\end{algorithm}

\subsection{Clustering and Model Selection}\label{sec:context-cluster}
The contextual GP can model the database performance in dynamic environments. 
However, it has  $O(n^3)$ complexity with $n$ observations. 
The cubical computation complexity limits the applicability with increasing observations in the cloud. 
To tackle this problem, we propose a clustering and model selection strategy based on the similarity of context features. 
The observations are clustered, and the number of observations in each cluster can be bounded under a constant number $P$. 
\sys fits multiple contextual GPs based on the clusters and learns a decision boundary for model selection.
Therefore the complexity is bounded by $O(P^3)$, which makes \sys can scale with increasing observations.

\begin{figure}
\centering
    \includegraphics[keepaspectratio=true,scale=0.55]{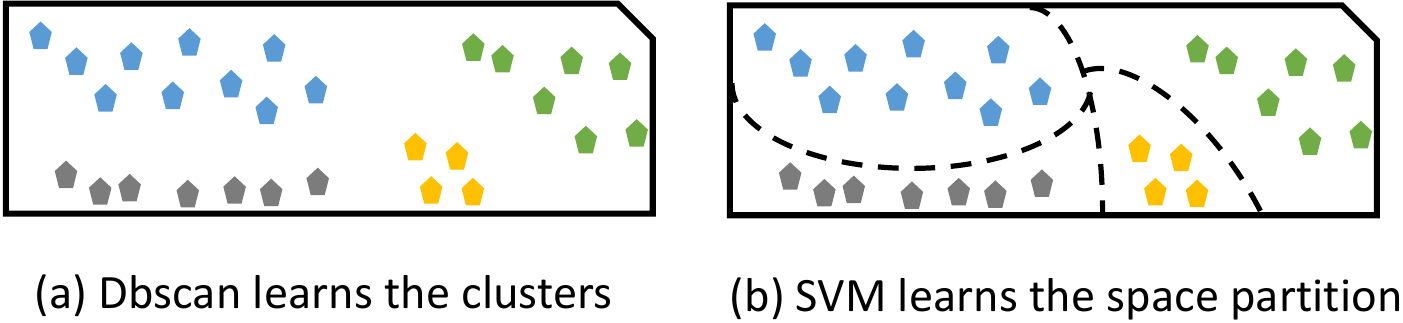}
\caption{Clustering and Model Selection: Each point denotes a sample and the learned boundary splits the context space.}
\label{fig:cluster}
\vspace{-1em}
\end{figure}

Algorithm \ref{alg:context-space} presents the procedure. 
We first perform the DBSCAN clustering algorithm ~\cite{DBLP:conf/kdd/EsterKSX96} based on context features $\{c_i\}_1^{t}$, obtaining a cluster label $l_i$ for each $c_i$ (Line \ref{alg:cluster}), as shown in Figure \ref{fig:cluster} (a).  
For each cluster, \sys fits a contextual GP model using its observations (Line \ref{alg:context-fit}).
To select a model for unseen contexts, \sys uses SVM to learn a non-linear decision boundary (Line \ref{alg:context-svm}, as shown in Figure \ref{fig:cluster} (b)). 
We choose SVM for its simplicity, ease of training, and the need for fewer samples to generalize well in practices~\cite{DBLP:conf/nips/WangFT20}. 
Besides improving scalability, such clustering excludes the observations with distant contexts from the training set for the GP model, preventing the ``negative transfer''~\cite{DBLP:conf/sigir/KrishnanDBYS20}.
 
Augmented observations are classified into a cluster based on their context and the learned boundary.
 However, the distribution of context features may shift as more observations are collected, and the previous clustering and boundary need to be re-learned periodically.  
\sys maintains the existing clustering and a simulated new clustering to determine whether to re-learn or not (Line \ref{alg:split}). 
The difference between the two kinds of clusterings indicates context shifts.
We use a mutual information-based score (MI) to quantify the difference.
MIs close to zero indicate two vastly dissimilar clusterings, while MIs close to one indicates the opposite. 
When the MI score is smaller than a threshold (we set 0.5 in the experiments),  the re-clustering is triggered, and the boundary and models are  updated based on the revised clusters  (Line \ref{alg:cluster}--\ref{alg:context-svm}). 


\section{Safe Configuration Recommendation}\label{sec:safe}
\sys aims to optimize the database performance while ensuring safety throughout the tuning process. 
Instead of optimizing the performance function over the global configuration space, \sys restricts its optimization in a subspace and gradually expands it towards the optimum.
The subspace restriction enables a fine discretization to generalize safety and mitigates the over-exploring nature of BO. 
This section introduces how \sys adapts the subspace, forms a safety set within the subspace, and selects a safe and promising configuration.

\subsection{Subspace Adaptation}\label{sec:safe-update}

To expand towards the global optimum, the configuration subspace is adjusted iteratively and alternated between hypercube and line regions, as shown in  Algorithm \ref{alg:subspace}.

The hypercube region 
$\{\theta \ \big| \left\|\theta-\theta_{best}\right\|_2 \leq R_n\}\cap \Theta$ is defined with a region center $\theta_{best}$ and a radius $R_m$. The region center $\theta_{best}$ is set to the best configuration observed so far.
The radius $R_n$ controls the optimization space. 
If $R_n$ is large enough for the hypercube region to contain the whole configuration space, this will be equivalent to running standard global BO, while a small $R_n$ may slow down the exploration. 
At initialization, $R_n$ is set to a base value (e.g., $5\%$ ranges of each dimension). 
It is typical behavior to shrink $R_n$ after  consecutive ``failures'' and expand it after  consecutive ``successes'' ~\cite{DBLP:journals/cj/NelderM65}. 
We define a ``success'' as recommending a configuration better than the previous one and a ``failure'' as a recommendation that does not. 
 The details about adjusting $R_n$ is illustrated in Line \ref{alg:r0}--\ref{alg:r1}. 

The line region is a one-dimensional affine subspace $\{\theta_{best} + \alpha d : \alpha \in \R | \}\cap \Theta$, defined with the offset  $\theta_{best}$ and a direction $d \in \R^{m}$.  
It is proved that the optimization with line regions can converge globally ~\cite{DBLP:conf/icml/KirschnerMHI019}. 
The direction of the line region determines the trace of optimization. 
We implement two strategies to generate the directions: random direction (increasing the exploration) and
important direction (aligned with the important configuration knob, increasing exploitation).  Appendix contains the details.
 
The subspace is first initialized as a hypercube region to restrict the optimization near the initial safety set (Lines~\ref{alg:init1}--\ref{alg:init2}). 
The hypercube region encourages optimization densely in the interior but may lead to an over-exploitation, even getting trapped in a local optimum \cite{DBLP:journals/corr/abs-2101-05147}. 
\sys switches to line region to control the trade-off between exploitation and safe exploration (Lines~\ref{alg:alter1}--\ref{alg:alter2}).
The alternation is triggered by a switching rule (Lines ~\ref{alg:stop1} and \ref{alg:stop2}): when no unevaluated safe configuration exits in $\Theta_m$ or a certain number of consecutive failures to recommend better configurations, \sys switches to another type of subspace. 
The update of $\theta_{best}$ moves the subspace towards the optimum.

\begin{algorithm}[t]
\DontPrintSemicolon
\KwIn{Current subspace $\Theta_{n}$, Current best config. $\theta_{best}$, \\
\hspace{0.93cm} Success counter $\Upsilon_{succ}$, Failure counter $\Upsilon_{fail}$}
\Parameter{Consecutive success threshold $\eta_{succ}$, Consecutive failure threshold $\eta_{fail}$}
\KwOut{Adapted configuration subspace}
\caption{Subspace Adaptation}\label{alg:subspace}
\If{$\Theta_n$ is None\label{alg:init1}}
{
$\Theta_n= \{\theta \ \big| \left\|\theta-\theta_{best}\right\|_2 \leq R_n\}\cap \Theta$. \;}\label{alg:init2}

\If{$\Theta_n$ is hypercube region\label{alg:alter1}}
{
\If{$ \Upsilon_{succ}$ > $\eta_{succ}$ \label{alg:r0}}
{
$R_n = min(R_{max}, 2R_{n})$, $ \Upsilon_{succ}=0$, $ \Upsilon_{fail}=0$.\;
}

\If{$ \Upsilon_{fail}$ > $\eta_{fail}$}
{
$R_n =R_n/2 $, $ \Upsilon_{fail}=0$, $\Upsilon_{succ}=0$.\;
}\label{alg:r1}

\If {Switching\_Rule()\label{alg:stop1}}
{

$d = generate\_direction()$.\;\label{alg:direction1}
$\Theta_n= \{\theta_{best} + \alpha d : \alpha \in \R | \}\cap \Theta$.
}\label{alg:direction2}
}

\If{$\Theta_n$ is line region}
{
\If {Switching\_Rule()\label{alg:stop2}}
{

$\Theta_n= \{\theta \ \big| \left\|\theta-\theta_{best}\right\|_2 \leq R_n\}\cap \Theta$.\;\label{alg:alter2}
}
}
\Return{$\Theta_n$}.\;
\end{algorithm}

\subsection{Safety Assessment}\label{sec:safe-assess}

\sys discretizes the adapted subspace to build a candidate set.
Then the safety of each candidate is assessed based on the confidence bounds of the contextual GP (black-box knowledge) and the existing domain knowledge (white-box knowledge).

\subsubsection{Black-Box Knowledge} 
 Given context $c$,  \sys utilizes the confidence bounds of the selected contextual GP model $m^n$  to access the safety of  $\theta$:
\begin{equation} 
\begin{aligned} 
l_{n}(\theta, c) &=   \mu_{n}(\theta, c) - \beta\sigma_{n}(\theta, c), \\ 
u_{n}(\theta, c) &=   \mu_{n}(\theta, c) + \beta\sigma_{n}(\theta, c), 
\end{aligned}
\end{equation}
where $l_n$ and $u_n$ are the lower and upper bound predictions, with $\mu_n$ and $\sigma_n$ from  Equation \ref{equ:gp}. 
The parameter $\beta$ controls the tightness of the confidence bounds, and we set its value following the study of Srinivas et al.~\cite{DBLP:conf/icml/SrinivasKKS10}.
The true function value $f$ falls into the confidence interval $[l_n(\theta, c),u_n(\theta, c)]$ with a high probability. 
We can determine safe configurations with $l_n(\theta, c)>\tau$; that is, configurations with worst-case performance still above the safety threshold $\tau$.
We restrict the performance modeling in a local subspace containing previously evaluated configurations, and the local modeling can be trusted more than global modeling.
This is also the rationale behind the trust-region methods from stochastic optimization ~\cite{Yuan_areview}.

\subsubsection{White-Box Knowledge} \label{sec:white}
Although the relationship between configurations and database performances is complex. 
Domain knowledge does exist to dismiss bad configurations. 
For example, the total buffer size can not exceed the physical memory capacity of the deployed machine.
Experienced DBA tunes the databases based on domain knowledge, and some database tuning tools also use heuristics to give tuning suggestions \cite{DBLP:conf/cidr/DiasRSVW05, DBLP:conf/vldb/WeikumMHZ02, DBLP:conf/fskd/WeiDH14,MySQLTuner}.
Such white-box tuning provides intuitive suggestions and could serve as a warm starting and space pruning component for ML-based tuning.
When forming the final safety set, \sys consults the white box and dismisses the unsafe or unpromising configurations deviated from the white box's suggestions. 
We implement \sys's white-box assistant using MysqlTuner~\cite{MySQLTuner}.
MysqlTuner examines the DBMS metrics and uses static heuristics to suggest setting ranges for configurations.
The ranges are generated based on rules (e.g., setting key buffer size larger than the total MyISAM indexes size or increasing the join buffer size if \#joins without indexes per day is larger than 250).
\sys removes the configurations not satisfying the white box's suggestions from its safety set.
If the white box suggests a specific configuration instead of the ranges, \sys filters the configurations far away from the suggested one.
Other white-box tuning tools or hand-crafted rules can coexist with  \sys if they provide suggestions for \sys's tuning knobs.
Instead of directly applying, their suggestions can be used as white-box assistants for \sys.

However, the white box does not evolve according to the feedback, causing the trap in local optimum.
The rules could even be inappropriate, excluding the optimal configurations from $S_n$. 
This happens only when the white-box rule rejects a configuration while the black-box algorithm recommends this configuration (i.e., decision conflict).
To prevent this, \sys uses a relaxation strategy to relax inappropriate rules. 
\sys maintains a  conflict counter and a conflict-safe counter for each rule. 
When the decision conflict happens several times, reaching a threshold, \sys will ignore the rule and recommend the controversial configuration. 
Note that only one rule can be ignored in the recommendation to control the interdependence between different rules. 
After evaluation, if the configuration is safe, the conflict-safe counter will increase by one. 
When a rule's conflict-safe counter reaches a threshold, the rule will be relaxed (e.g., the configuration range given by the rule is enlarged).
The thresholds for each white-box rule can be set differently according to its credibility. 
A larger threshold leads to more trust in the white-box rule.

\subsection{Candidate Selection}\label{sec:safe-opt}

After the safety set is generated, \sys selects a configuration from the candidates in the safety set. 
Like all the black-box optimization, the selection should trade-off between two objectives: (1) exploitation, trying to localize the high-performance regions within the current safety set, (2) exploration, acquiring new knowledge, and trying to expand the current estimate of the safety set. 
We adopt Upper Confidence Bound (UCB)~\cite{DBLP:conf/icml/SrinivasKKS10} constrained to the safety set as a sampling criterion, shown in Equation \ref{equ:ucb}:
\begin{equation}
\theta_{max}= \mathop{\arg\max}_{\theta\in S_n}{\mu_{n}(\theta, c) + \beta\sigma_{n}(\theta, c)}\label{equ:ucb}.
\end{equation}
UCB selects the configuration at locations where the upper bound of its confidence interval is maximal. 
Repeatedly evaluating the system performance at configurations given by UCB improves the mean estimate of the underlying function and decreases the uncertainty at candidate configurations for the maximum. 
The global maximum is provably found eventually~\cite{DBLP:conf/icml/SrinivasKKS10}.

To expand the safe subspace explicitly, \sys also selects the safe configurations at the boundary of the safety set with the highest uncertainty since they are promising candidates for expanding the safety set. 
To unify the two sampling criteria, we adopt the epsilon-greedy policy ~\cite{DBLP:journals/corr/MnihKSGAWR13}, which selects the maximal UCB configuration with a probability of $1-\epsilon$ and the boundary point with a possibility of $\epsilon$.

\section{Evaluation}\label{sec:exp}

\noindent\textbf{Outline}:
We first compare \sys with the state-of-the-art database configuration tuning methods on dynamic environments in \autoref{sec:exp-dyn}. 
The comparison is conducted under three settings: (1) \textit{dynamic queries composition}, where the queries composition of the workload is constantly changing, (2) \textit{transactional-analytical}, simulating a daily transactional-analytical workload cycle by alternating the execution of OLTP and OLAP workloads, and (3) \textit{real-world workload}.
In \autoref{sec:exp-overhead}, we analyze the overhead of the compared methods. 
In \autoref{sec:exp-case}, a case study is presented.
In \autoref{sec:analysis}, we conduct ablation studies to evaluate the effectiveness of \sys's design, including the context space design and the safe exploration strategy. 
We also validate the robustness of \sys on different initial safety sets and interval sizes. 
In addition, we evaluate the baselines under \textit{static workload} setting to analyze the search efficiency in \autoref{sec:eosw}.

\noindent\textbf{Baselines}: The baselines are explained below:
\begin{itemize}[leftmargin=*]
\item \underline{\textit{DBA Default}} is the configuration provided by experienced DBAs.
\item \underline{\textit{}{\sys}}  is our safe and contextual tuner. We implement \sys's prediction model using  GPy library~\cite{gpy}.

\item \underline{\textit{BO}} is a Bayesian Optimization approach, widely used in database configuration tuning ~\cite{DBLP:journals/pvldb/DuanTB09, DBLP:conf/sigmod/AkenPGZ17, DBLP:conf/hotstorage/KanellisAV20,DBLP:journals/pvldb/AkenYBFZBP21}. We use similar design with OtterTune~\cite{DBLP:journals/pvldb/DuanTB09}: Gaussian process as surrogate model and EI  as acquisition function. We also implemented BO via GPy library.
\item \underline{\textit{DDPG}}   is a  reinforcement learning agent which is used to tune the database configuration ~\cite{DBLP:conf/sigmod/ZhangLZLXCXWCLR19}. 
The agent inputs internal metrics of the DBMS and outputs proper configurations.
The DDPG algorithm is implemented using PyTorch library~\cite{pytorch} with its neural network architecture borrowed from CDBTune~\cite{DBLP:conf/sigmod/ZhangLZLXCXWCLR19}.

\item \underline{\textit{QTune}} is a query-aware tuner that supports three tuning granularities~\cite{DBLP:journals/pvldb/LiZLG19}.
We adopt its workload-level tuning.
It embeds workload features to predict
internal metrics via a pre-trained model, while CDBTune uses the measured internal metrics.

\item \underline{\textit{ResTune}} adopts  constrained Bayesian Optimization to minimize resource utilization with SLA constraints~\cite{DBLP:conf/sigmod/ZhangWCJT0Z021}.
It uses an ensemble
framework (i.e., RGPE) to transfer historical knowledge from observations of source workloads.
We modify ResTune to maximize database performance with the same safety constraints as \sys.
To adopt RGPE in online tuning, we cluster every 25 observations as one source workload.

\item \underline{\textit{MysqlTuner}} is a MySQL tuning tool that examines DBMS metrics and uses static heuristics to suggest configurations~\cite{MySQLTuner}. It is also the white-box assistant that \sys consults. 
\end{itemize}

\noindent\textbf{Workloads}:
We use three workloads with different characteristics from well-known benchmarks and a real-world workload. The three workloads are TPC-C, Twitter from OLTP-Bench~\cite{DBLP:journals/pvldb/DifallahPCC13} and JOB~\cite{DBLP:journals/pvldb/LeisGMBK015}. 
TPC-C is a traditional OLTP benchmark characterized by write-heavy transactions with complex relations. 
Twitter is extracted from web-based applications, characterized by heavily skewed many-to-many relationships and non-uniform access. 
JOB is an analytical workload with 113 multi-join queries, characterized by realistic and complex joins~\cite{DBLP:journals/pvldb/LeisGMBK015}.
We load about 29 GB data for Twitter, 18 GB for TPC-C, 9 GB for JOB and use unlimited arrival rates for OLTP workloads to fully evaluate the benefits from tuning, as ~\cite{DBLP:journals/pvldb/DuanTB09, DBLP:conf/sigmod/AkenPGZ17, DBLP:conf/hotstorage/KanellisAV20,DBLP:conf/sigmod/ZhangLZLXCXWCLR19, DBLP:journals/pvldb/LiZLG19} did. 
The real-world workload comes from a database monitoring service. 
In our experiments, we use the workload from 10:00 to 16:00 on September 2nd, 2021. 
And the read-write ratio per minute varies from 3:1 \textasciitilde 74:1 in this period.

\noindent\textbf{Setup}:
We use version 5.7 of RDS MySQL. 
We tune 40 dynamic configuration knobs without restarting the database since the restart is not acceptable for online databases. 
The 40 configuration knobs are chosen based on their importance by DBAs. 
The experiments run on a cloud instance with 8 vCPU and 16GB RAM. 
The interval size is set to 3 minutes.
We use the DBA default configuration as the initial safety set and its performance as the safety threshold, which are also added to the training set of other baselines for fairness.

 \begin{figure}[t]
\centering
    \includegraphics{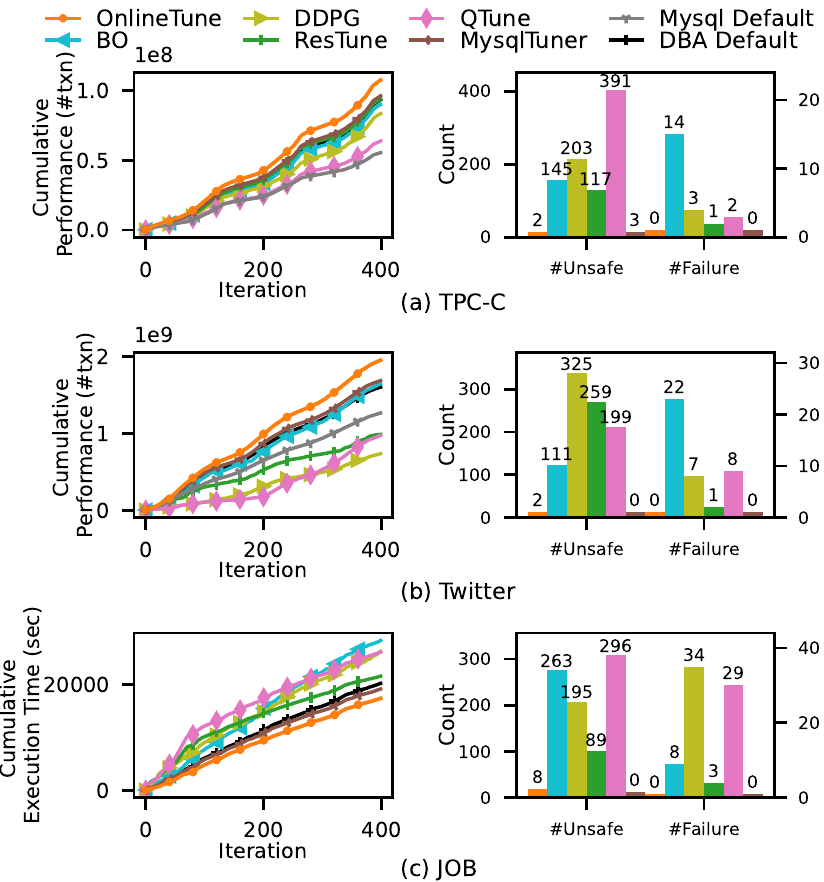}
\caption{Cumulative performance (for TPC-C and Twitter, the higher is the better; for JOB, the lower is the better) and safety statistics when tuning dynamic workloads.}
\label{fig:dynamic-improvement}
\end{figure}

\begin{figure*}
\centering
    \includegraphics{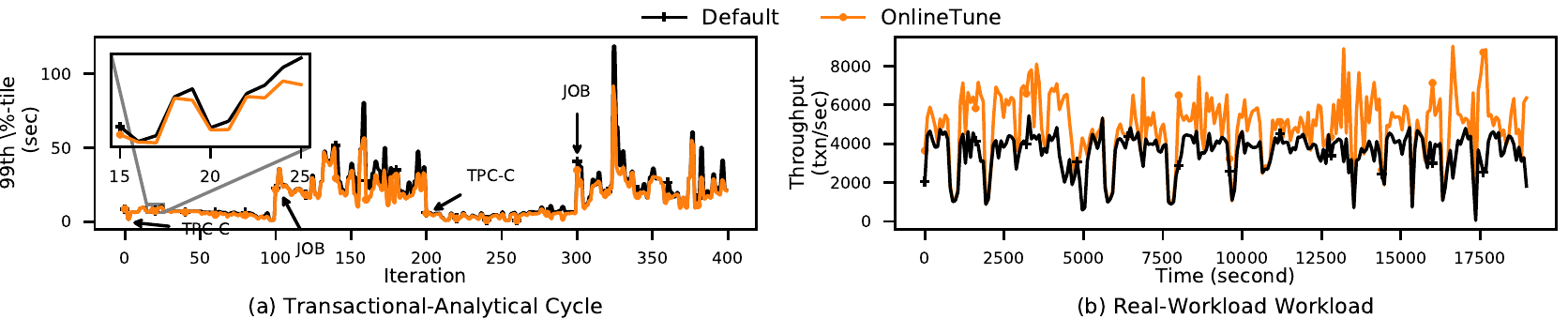}
\caption{Iterative Performance on OLTP-OLAP circle and real-world workload.}
\label{fig:real-performance}

\end{figure*}
 
  \begin{figure*}[t]
\centering
    \includegraphics{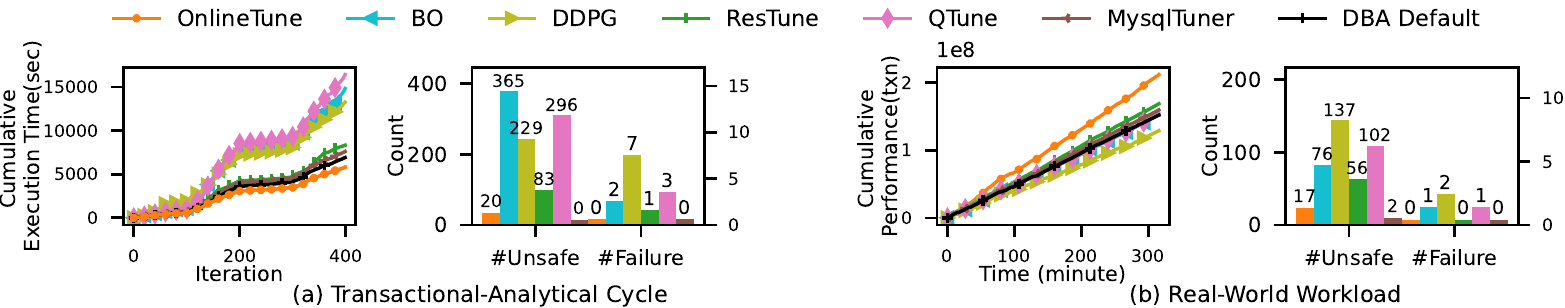}
\caption{Cumulative performance and safety statistics on OLTP-OLAP circle and  real-world workload.}
\label{fig:real-improvment}

\end{figure*}

\begin{figure}[t]
\centering
    \includegraphics{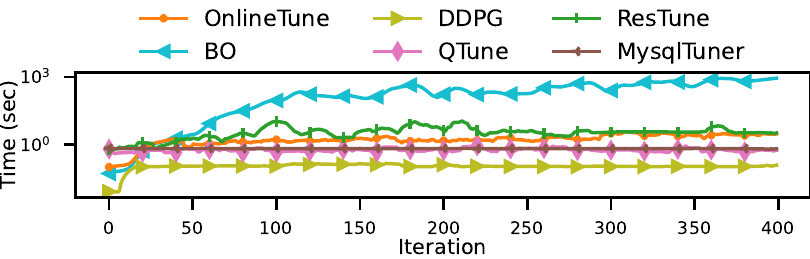}
\caption{Computation time of different approaches.}
\label{fig:time}
\end{figure}

\noindent\textbf{Metrics}: 
We evaluate the baselines from two perspectives: safety and cumulative performance during tuning.
For safety, we count the number of unsafe configuration recommendations (\#Unsafe) and the number of system failures (\#Failure) within the tuning period. 
For the cumulative performance, it is measured by the number of transactions (\#txn) processed by the database during tuning for OLTP workload and the sum of execution time for OLAP workload.

\subsection{Evaluation on Dynamic Workload}\label{sec:exp-dyn}
We evaluate the baselines tuning online database with constructed dynamic workloads and real-workload workload with dynamic query arrival rates.
We intend to answer two questions: (1) \textit{Can they recommend the configuration adaptively with the dynamic workload?} (2) \textit{Can they 
reliably respect the safety requirement?}

\subsubsection{Evaluation on workloads with dynamic query compositions}\label{sec:exp-dynamic}
 
To simulate the dynamicity, we construct workload with dynamic query compositions.
For TPC-C and Twitter, we vary transaction weights via OLTP-Bench~\cite{DBLP:journals/pvldb/DifallahPCC13}. 
The weights are sampled from a normal distribution with a sine function of iterations as mean and a $10\%$ standard deviation.
For JOB, we execute ten queries per iteration, and five out of them are re-sampled. 
If the execution time exceeds the interval size, we kill the queries.
Since TPC-C is a write-heavy workload, its underlying data is also changing (e.g., its data size change from  18 GB to  48 GB during tuning). 

\begin{figure*}[htbp]
\centering
\begin{minipage}[t]{0.3\textwidth}
\centering
\includegraphics[width=5cm]{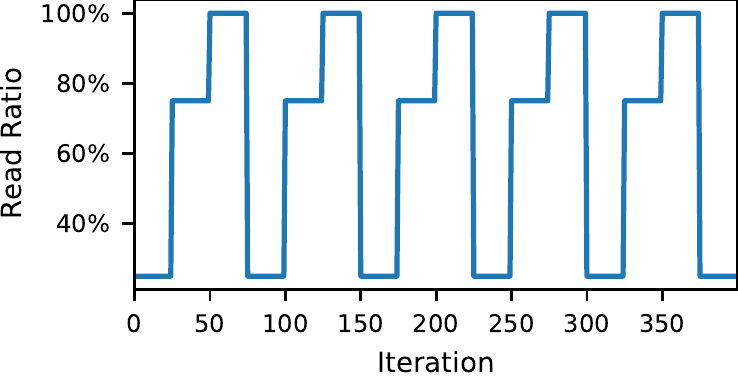}
\caption{Pattern of YCSB workload.}\label{fig:workload-trace}
\end{minipage}
\begin{minipage}[t]{0.6\textwidth}
\centering
\includegraphics[width=10cm]{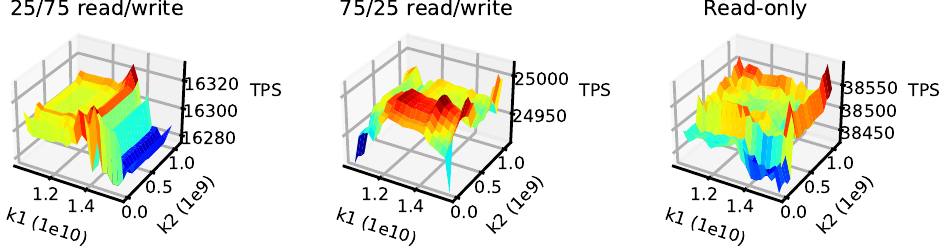}
\caption{Throughput as a function of configurations: k1 denotes sort\_buffer\_pool\_size and k2 denotes max\_heap\_table\_size.}\label{fig:workload-surface}
\end{minipage}
\end{figure*}

  \begin{figure*}[t]
\centering
    \includegraphics{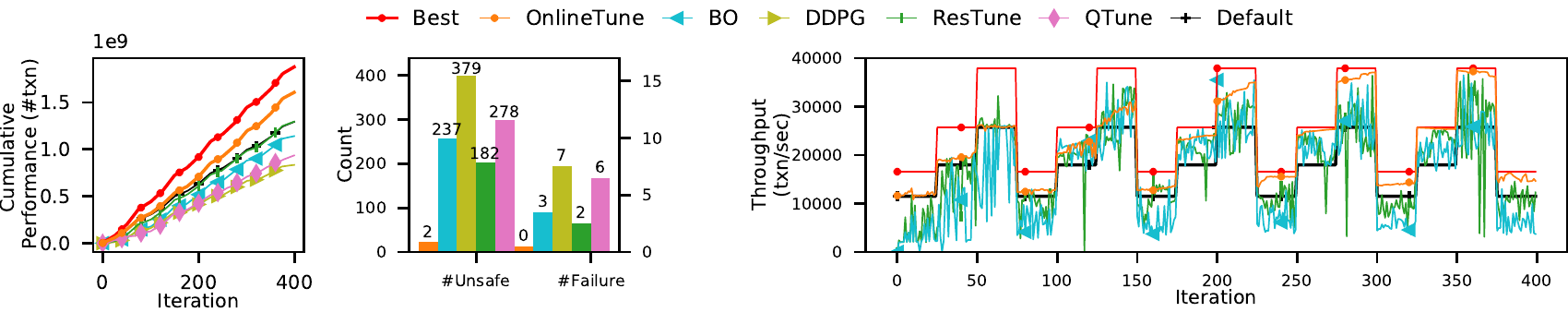}
\caption{Results on YCSB workload: The left figure shows the cumulative result and the right shows the iterative performance. 
For clarity of presentation, we only show the iterative performances of the top 3 baselines (\sys, ResTune and BO).}\label{fig:case-result}
\label{fig:dynamic-instance}

\end{figure*}

\vspace{0.3em}
\noindent\textbf{\sys finds the workload-specific configuration.}
As shown in Figure~\ref{fig:dynamic-improvement},  \sys  achieves  54.3\% \textasciitilde 93.8\% improvement on  cumulative performance than the MySQL default, and 16.2\% \textasciitilde 21.9\% improvement than the DBA default. 
The DBA default is expected to have a relatively robust performance across workloads.
\sys applies better configuration, illustrating its adaptability in dynamic environments.
In addition, \sys achieves 14.4\%\textasciitilde 165.3\% improvement on cumulative performance than existing offline approaches.
The reasons are two-fold.
First, the offline methods struggle to handle dynamic environments.
BO uses observations $\{\theta_i, y_i\}_1^{t}$ to fit the GP model, ignoring dynamic environmental factors. 
When workload drifts, BO fails to learn suitable configurations.
ResTune adopts an ensemble GP that assigns different weights to its base models but its base models still ignore the dynamic factors.
DDPG and QTune fine-tune the models when environments change, but fine-tuning a neural network needs lots of training samples, which is inefficient for online tuning.
Second, the offline methods over-explore configuration spaces. 
Evaluating unsafe configurations is a part of learning for offline methods.
Although ResTune aims at finding the configurations satisfying the constraints, it still needs to evaluate and learn the unsafe area.
Compared with the white-box method (MysqlTuner), \sys achieves 10.1\%\textasciitilde 19.7\% improvement.
Although MysqlTuner does not have the over-exploration issue, it relies on heuristic rules and  traps in local optimum.

\vspace{0.3em}
\noindent\textbf{ \sys reliably respects the safety requirement when tuning the online database.} 
Recommending safe configurations is non-trivial when tuning online databases since the workload changes may cause the shifting of safe configurations. 
As shown in Figure \ref{fig:dynamic-improvement}, none of the system failures occur when applying \sys, while offline tuning methods cause several system failures. 
The offline methods have 22.2\% \textasciitilde 97.8\% unsafe configuration recommendations within the 400 tuning intervals. 
Compared to them, \sys reduces 91.0\%\textasciitilde99.5\% unsafe recommendations.
This is contributed to the safe exploration strategy of \sys, which we analyze in detail with an ablation study in \autoref{sec:ablation-safe}.

\subsubsection{Evaluation on Transactional-Analytical Cycle}
We simulate a daily transactional-analytical workload cycle by alternating dynamic TPC-C and JOB workloads. 
We use 99\% latency as the optimization objective. 
The workload characterization of TPC-C (OLTP, write-heavy) and JOB (OLAP, complex joins) are significantly different. 
Respecting safety constraints is rather tricky in such a scenario.
Figure \ref{fig:real-performance} (a) shows the performance of \sys and the DBA default. 
The workload starts as TPC-C and repeatedly alternates with JOB every 100 iterations.  
\sys gradually finds configurations better than the DBA default, as shown in the zoom-in plot. 
When switching to JOB, \sys takes some iterations to explore the configuration space. 
Meanwhile, several unsafe configurations with latency slightly larger than the safety threshold occur. 
Then, \sys finds suitable configuration adaptively (e.g., larger sort buffer size for JOB workload).
When switching from TPC-C to JOB again, \sys selects the surrogate model fitted with previous observations for JOB and recommends suitable configurations more quickly. 
\sys achieves performance than the DBA default while respecting the safety requirement, while other approaches fail, as shown in Figure \ref{fig:real-improvment} (a).

  \begin{figure*}[t]
\centering
    \includegraphics{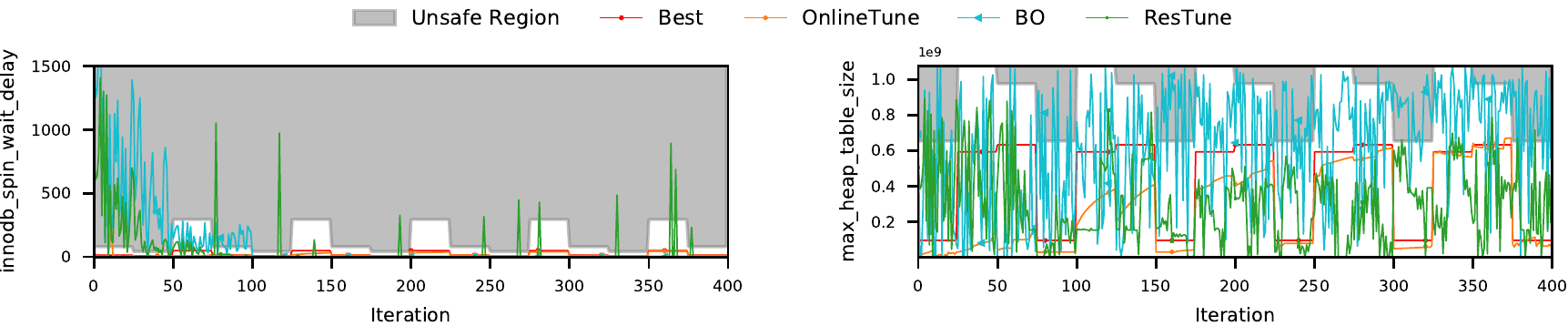}
\caption{Configurations applied by \sys, ResTune and BO on YCSB workload: We present the top 2 important knobs.}\label{fig:case-knob}
\label{fig:dynamic-instance}

\end{figure*}

\begin{figure*}[t]
\centering
    \includegraphics{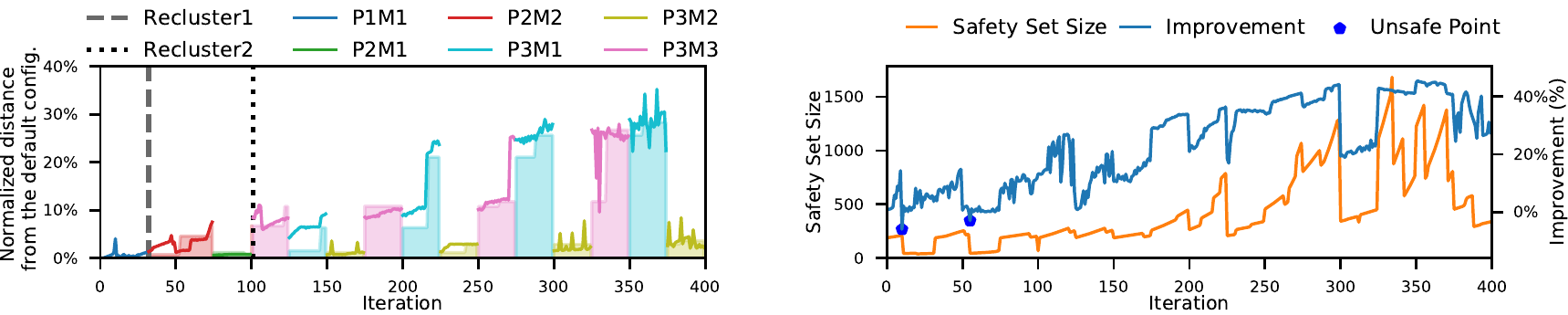}
\caption{Visualization of \sys. 
In the left figure, different colors denote different models selected over iterations. 
The lines denote the distance between the recommended configuration and the default, and the dash denotes the distance between subspace centers and the default.
The right figure presents the size of safety set estimated by \sys in its restricted subspace.
For reference, we also plot the performance improvement over iterations.}\label{fig:case-process}
\end{figure*}

\subsubsection{Evaluation on Real-World Workload}
We evaluate on a workload from the real-workload application.
Figure \ref{fig:real-performance} (b) shows the tuning process of \sys and Figure \ref{fig:real-improvment} (b) presents the performances of all the baselines. 
We observe that \sys achieves 39.4\% improvement on cumulative improvement compared to the DBA default and 25.5\% \textasciitilde 64.2\% improvement compared to offline methods. 
Although \sys applies several unsafe configurations at the beginning, their performances are within 10\% deviation of the default performance as shown in Figure \ref{fig:real-performance} (b).

\subsubsection{Algorithm Overhead}\label{sec:exp-overhead}
Figure~\ref{fig:time} shows the computation time when tuning JOB workload. 
For BO, DDPG and ResTune, the computation time consists of (1) statistics collection, (2) model fitting, and (3) model probe. 
For \sys and QTune, the time also includes featurization. 
\sys's computation time is slightly larger than BO at the beginning. 
However, BO's overhead increases exponentially over tuning iteration, as GP suffers from cubic complexity on sample number. 
Instead, \sys's computation time is within 3.79 seconds due to its clustering strategy. 

In addition to the computation time, we also analyze the impact of \sys's tuning overhead.
\sys's featurization module is deployed in the database instance, while the other parts are deployed in the backend tuning severs whose overhead does not influence the database instance.
Therefore, we focus on the impact of featurization.
This module takes about 57.7 ms per iteration on average.
To measure its impact, we keep the configuration unchanged and compare the resource usages and database performance.
We observe that the average CPU usage with featurization is 77.28\%, and the one without featurization is 77.27\%.
The increase is negligible since the featurization time only accounts for a slight proportion of a tuning interval (180 s).
Its impact on database performance can also be overlooked due to marginal resource consumption.
For evaluation, we keep configurations unchanged to observe the impact.
In practice, the database performance will be improved when running \sys, as evaluated above.



\subsection{Case Study}\label{sec:exp-case}

To further investigate \sys's tuning performance, we conduct a case study tuning five knobs.
We construct a workload trace using YCSB with different read/write transaction compositions, as shown in Figure \ref{fig:workload-trace}.
The joint context-configuration space is smaller than the other experiments in the paper. 
Therefore, we can use extensive evaluations to explore the space and obtain the best configurations (denoted as the Best) and the unsafe areas for each workload composition.
As shown in  Figure \ref{fig:workload-surface}, we observe that there are interactions between knobs. 
And, regular patterns exist among workload compositions, e.g.,  when buffer pool size (k1) is large and  heap table size (k2) is small, the throughput is relatively low.
However, the overall effects of knobs are quite different among the workloads, e.g., distinct optimal positions.
Thus, it is necessary to consider the workload dynamicity when tuning online.

Figure \ref{fig:case-result} presents the tuning result, which aligns with the evaluation in ~\autoref{sec:exp-dyn}.
As shown in the iterative plot, the distances between \sys and the Best gradually decrease as \sys safely finds configurations near the optimum.
The iterative performances of the offline baselines have large fluctuations due to their trials and errors.
In Figure \ref{fig:case-knob}, we focus on the configuration values of the top-2 important knobs. 
The unsafe region is approximated by excluding the knob's values from all the safe configurations. 
Note that if a configuration is assigned with a knob value in the unsafe region, the configuration is unsafe, but not vice versa. 
This is due to knobs interaction may cause unsafe.
We observe that the optimal configurations (the Best) are not portable across workloads.
\sys applies configurations adaptive to the dynamic workload, which safely proceeds towards the optimum. 
The other baselines explore the  space aggressively in the first 50 iterations and always have a chance to explore the unsafe region.

Figure \ref{fig:case-process} visualizes the working process of \sys's modules. 
From the left figure, we make two observations.
First, \sys can select the corresponding model for the observed context.
Second, the subspace is initially centered around the default configuration and moves towards the optimum. 
The configuration subspaces maintained by each model gradually move far from the default, safely exploring the configuration space.
From the right figure, we observe that the size of safety set is updated with the augmented observations.
If \sys evaluates unsafe configuration,  its safety estimation will be immediately tightened, and \sys will recommend conservative configurations near the evaluated-best ones, avoiding successive regression.

\subsection{Analysis of \sys}\label{sec:analysis}
We carefully design \sys with contextual modeling and safe exploration strategy to explore configurations safely in dynamic environments. 
We evaluate the corresponding designs via ablation study and validate the robustness of \sys.

\begin{figure}[t]
\centering
    \includegraphics{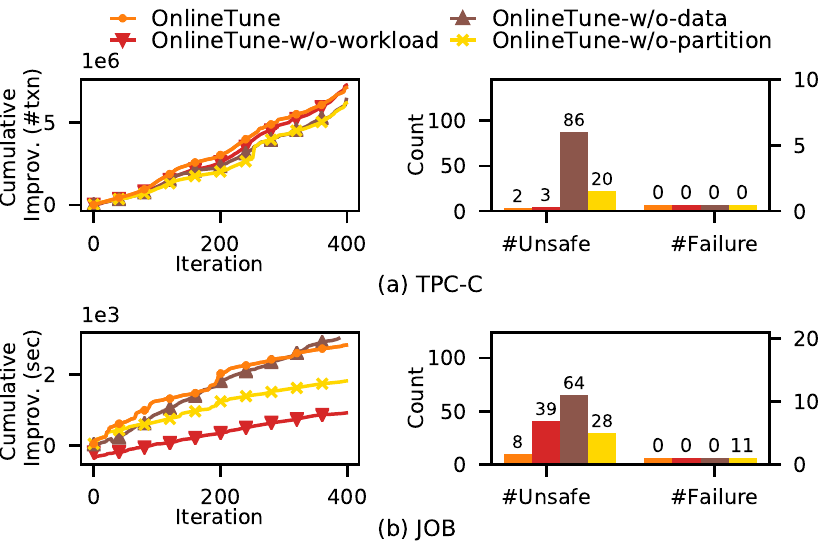}
\caption{Ablation study  on context space design.}
\label{fig:ablation-contex-improvement}

\end{figure}

  \begin{figure}[t]
\centering
    \includegraphics{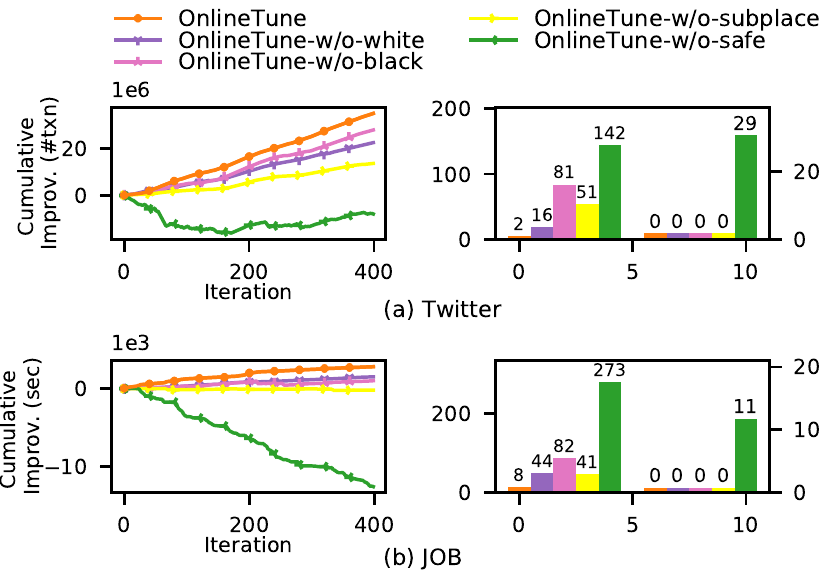}
\caption{Ablation study  on safe exploration.}
\label{fig:ablation-safe-improvement}

\end{figure}

\subsubsection{Ablation Study on Contextual Modeling}
We remove certain components of contextual modeling (featurization of workload and data changes, clustering, and model selection) in \sys to understand their contribution to the overall system.  
We compare (1) \sys, (2) \sys-w/o-workload, removing workload feature from context, (3) \sys-w/o-data, removing optimizer statistics (underlying data feature) from context, (4)   \sys-w/o-clustering, removing clustering and model selection design.
The experiments are conducted in dynamic TPC-C and JOB with the same setting as in \autoref{sec:exp-dynamic}.
Figure \ref{fig:ablation-contex-improvement} shows the improvement in cumulative performance against the DBA default and safety statistic of those baselines.
In the following evaluation, instead of cumulative performance, we use cumulative improvement that shows the benefit of online tuning more clearly.
For the read-only workload JOB where no data changes occur, \sys-w/o-data exceeds \sys slightly because the dimension of context feature decreases with no information loss, which means that modeling over contexts becomes easier. 
However, when data change occurs as in TPC-C workload, \sys outperforms the other baselines since its context feature  gives a comprehensive abstraction of the dynamic environment. 
Since the clustering and model selection strategy can prevent ``negative transfer'', \sys outperforms \sys-w/o-clustering in all the cases.

\begin{figure}[t]
\centering
    \includegraphics{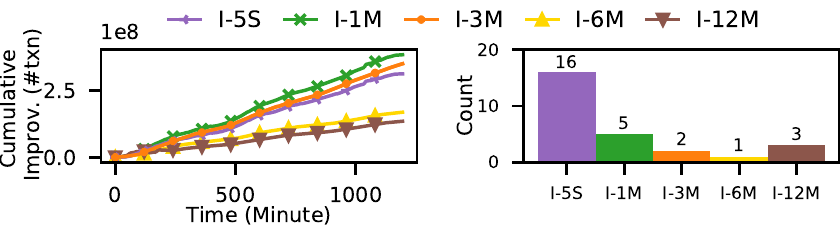}
\caption{Tuning Twitter with different interval sizes.}
\label{fig:interval}
\end{figure}

  \begin{figure}[t]
\centering
    \includegraphics{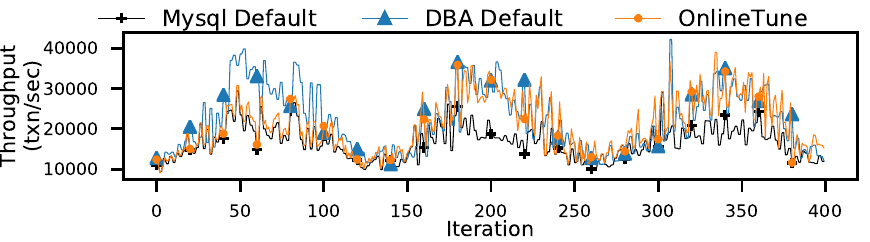}
\caption{Starting from MySQL Default.}
\label{fig:vary-start}

\end{figure}
 
 \begin{table*}
\caption{Statistics on static workloads: For each workload, bold face indicates the best value. Search Step denotes the iteration needed to find a configuration within $10\%$ of the estimated optimum, $\backslash$ denotes such configuration is not found.} 
\label{tab:fix}
\begin{adjustbox}{max width=\textwidth}
\begin{tabular}{@{}ccccccccccccc@{}}
\toprule
                             & \multicolumn{2}{c}{\textbf{\sys}}              & \multicolumn{2}{c}{\textbf{BO}}                                   & \multicolumn{2}{c}{\textbf{DDPG}}                            & \multicolumn{2}{c}{\textbf{ResTune}}                              & \multicolumn{2}{c}{\textbf{QTune}}                           & \multicolumn{2}{c}{\textbf{MysqlTuner}} \\
\textbf{Workload}            & Max Improv.      & Search Step                       & Max Improv.      & Search Step                           & Max Improv. & Search Step                           & Max Improv.      & Search Step                           & Max Improv. & Search Step                           & Max Improv. & Search Step      \\ \midrule
\multicolumn{1}{c|}{TPC-C}   & 17.03\%          & \multicolumn{1}{c|}{176}          & \textbf{19.99\%} & \multicolumn{1}{c|}{\textbf{74}}      & 16.66\%     & \multicolumn{1}{c|}{76}               & 12.03\%          & \multicolumn{1}{c|}{\textbackslash{}} & 12.02\%     & \multicolumn{1}{c|}{\textbackslash{}} & 13.44\%     & \textbackslash{} \\
\multicolumn{1}{c|}{Twitter} & \textbf{48.18\%} & \multicolumn{1}{c|}{129}          & 43.43\%          & \multicolumn{1}{c|}{158}              & 35.79\%     & \multicolumn{1}{c|}{\textbackslash{}} & 46.95\%          & \multicolumn{1}{c|}{\textbf{10}}      & 8.06\%      & \multicolumn{1}{c|}{\textbackslash{}} & 13.07\%     & \textbackslash{} \\
\multicolumn{1}{c|}{JOB}     & 11.67\%          & \multicolumn{1}{c|}{\textbf{141}} & 7.77\%           & \multicolumn{1}{c|}{\textbackslash{}} & 7.60\%      & \multicolumn{1}{c|}{\textbackslash{}} & \textbf{11.84\%} & \multicolumn{1}{c|}{155}              & 11.24\%     & \multicolumn{1}{c|}{168}              & 7.23\%      & \textbackslash{} \\ \bottomrule
\end{tabular}
\end{adjustbox}
\end{table*}

  \begin{figure*}
\centering
    \includegraphics{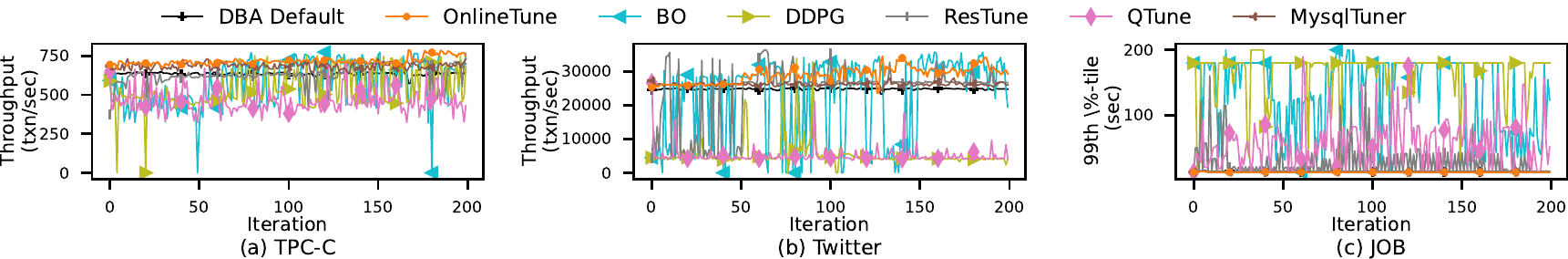}
 \caption{Iterative performance on static workloads: minimal throughput (0) and maximal latency (200) denote system failure.}
\label{fig:fix-performance}
\end{figure*}

\subsubsection{Ablation Study on Safe Exploration}\label{sec:ablation-safe}
To safely optimize the online database, \sys 
utilizes contextual GP (black-box knowledge) and domain knowledge (white-box knowledge)  and gradually expands the optimization subspace. 
To analyze their functionality, we compare: (1) \sys, (2) \sys-w/o-white, removing white-box safety assessment (3) \sys-w/o-black, removing black-box safety assessment, (4)\sys-w/o-subspace, optimizing in the whole configuration space, (5) \sys-w/o-safe, removing all the safety strategy, i.e., vanilla contextual BO. 
As shown in Figure \ref{fig:ablation-safe-improvement}, \sys beats other baselines in both performance  and safety degree. 
We make the following observations: 
(1) The black-box knowledge largely reduces the unsafe recommendations. 
The limited domain rules only cover a small subset of unsafe cases and fail to capture the complex and high-dimension relationship between configuration, environment, and database performance. 
Therefore, \sys-w/o-black's unsafe recommendations are much more than \sys-w/o-white's. 
(2) The white-box knowledge assists in filtering the unsafe configurations. 
We find that the unsafe recommendation in \sys-w/o-white is mainly caused by the knobs without intrinsic ordering, e.g., thread\_concurrency. 
Thread\_concurrency defines the maximum number of threads permitted inside InnoDB. 
But a value of 0 (the default value) is interpreted as infinite concurrency (no limit).  
The GP model depends on the natural ordering property and smoothness of space. 
When \sys-w/o-white expands the safe space, it is likely to try a value near zero, like one, casing performance downgrade due to the lack of computing resources. 
\sys has domain rules to filter the value of thread concurrency less than half of the number of virtual CPUs, preventing the unsafe case. 
(3) Optimizing over promising subspace instead of the whole configuration space can enhance safety and localize good configurations. 
\sys-w/o-subspace recommends more unsafe configurations than \sys.
As discussed, it is hard to generalize the safety of configurations in the whole space, and the prediction of GP is more trustworthy in a small trust region. 
Besides, the global optimization approaches (i.e., \sys-w/o-subspace, \sys-w/o-safe)  they over-explore boundaries of configuration spaces. 
(4) Without any safety designs, \sys-w/o-safe has the worst performance. 


\subsubsection{Varying   Interval Sizes}\label{sec:exp-interval}
 We run \sys under different interval sizes: 5 seconds, 1 minute, 3 minutes, 6 minutes, and 12 minutes, as shown in Figure \ref{fig:interval}.
 Within a certain range, a smaller interval leads to quicker adaptation.
 The reason is two-fold.
 First, OnlineTune could recommend fine-grained configurations suitable for dynamic workloads.
Second, \sys could collect more observations in a given time, leading to better tuning policies.
However, the interval size also determines the time for evaluating the database performance.
It cannot be too small to avoid the instability caused by performance fluctuation~\cite{DBLP:journals/pvldb/AkenYBFZBP21}.
As shown, tuning with a 5-second interval performs worse than the 1-minute one with more unsafe recommendations. 
And we have observed significant performance variance for 5-second intervals on a fixed configuration.

\vspace{-0.3em}

\subsubsection{Varying Initial Safety Set and Safety Threshold}\label{exp:safety-set}
In the above evaluation, we use the DBA default as the initial safety set and its performance as a safety threshold. 
However, there exists a question: \textit{Can \sys recommend suitable configurations with an 
inferior starting point?}
Therefore, we use the MySQL default configuration as the initial safety set and its performance as the safety threshold. 
As shown in Figure \ref{fig:vary-start}, the MySQL default's performance is worse than the DBA default's. 
One main difference is that the MySQL default sets the buffer pool size  128 MB while the DBA default sets the buffer pool size 13 GB.
When starting from the MySQL default, \sys applies safe configurations better than the MySQL default.
And it achieves a comparable performance to tuning with the DBA default as the starting point after about 150 iterations.

 \subsection{Evaluation on Static Workload}\label{sec:eosw}
The existing approaches work well to search for optimal configurations on  static workloads. 
This evaluation aims to assess the search efficiency of \sys with the safety constraints. 
Figure \ref{fig:fix-performance} presents the performance with statistics shown in Table \ref{tab:fix}.

\noindent\textbf{\sys reduces the unsafe recommendations significantly with search efficiency comparable to the state-of-the-art offline tuning methods.}
Offline methods are designed to search for the  optimum without safety considerations. 
In terms of efficiency, \sys is comparable to BO, ResTune, and better than DDPG, QTune in all the cases. 
\sys is very unlikely to apply unsafe configurations, while the offline methods violate safety constraints considerably. 
Although \sys's safety consideration may make the exploration slower, adaptively restricting the search space could localize good solutions, which improves the convergence, especially in JOB's case.

\section{Conclusion}\label{sec:conclusion}
We introduce \sys, an online tuning system that 
is aware of the dynamic environments and optimizes the database safely. 
\sys featurizes the dynamic environmental factors as context feature and leverages Contextual Bayesian Optimization to optimize the context-configuration joint space. 
We propose a safe exploration strategy, greatly enhancing the safety of online tuning. 
As future extensions, we plan to investigate combining \sys with offline tuning.
The offline process could explore more configurations on replicas of the target DBMS, replaying the historical workloads to collect observations for online tuning.
Therefore, \sys could exploit the promising configuration space in the online phase, thus responding to the dynamic environment more quickly.
In addition, \sys could pause online configuring after applying suitable configurations.
This could be achieved by a stopping and triggering mechanism.
We could keep \sys's workflow at each iteration (including context featurization and acquisition value calculation for candidate points) but not change the database configurations until more promising candidates appear.
For example, we can measure whether more promising candidates exist by calculating the Expected Improvement (EI) value against the applied configuration.
The configuring is triggered when candidates with EI values larger than a threshold exist, indicating that the context changes lead to the need for re-configuring.

\begin{acks}
This work is supported by National Natural Science Foundation of China (NSFC)(No. 61832001), Alibaba Group through Alibaba Innovative Research Program and National Key Research, and the Beijing Academy of Artificial Intelligence. 
Bin Cui is the corresponding author.
\end{acks}

\bibliographystyle{ACM-Reference-Format}
\balance
\bibliography{main}

\newpage
\appendix
\setcounter{table}{0}
\setcounter{figure}{0}
\setcounter{section}{0}
\renewcommand{\thetable}{A\arabic{table}}
\renewcommand\thefigure{A\arabic{figure}}
\renewcommand\thesection{A\arabic{section}}

\noindent\textbf{\Huge{Appendix}}
\vspace{1em}

\section*{Outline}
This supplemental material is organized as follows:

\noindent\textbf{A.1} More details about \sys Architecture.

\noindent\textbf{A.2} High-Level Algorithm.

\noindent\textbf{A.3} Details about subspace adaptation.


\section{More details about \sys Architecture}\label{appendix}\label{sec:app-sys}
\begin{figure}[h]
\centering
    \includegraphics{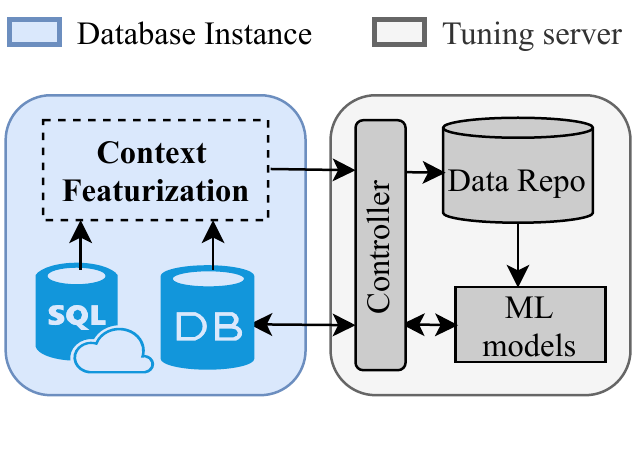}
\caption{\sys Architecture.}
\label{fig:sys}
\end{figure}

\noindent\textbf{System Architecture.}
Figure \ref{fig:sys} presents the architecture of \sys. 
The left part shows the online database system in the cloud environment that runs ever-changing workloads. 
The right part represents the \sys server deployed in the backend tuning cluster.
\sys maintains a data repository that stores the historical observations  $\{<c_i, \theta_{i}, y_i>\}_1^{t}$ from the previous tuning iterations, which could be initially empty.
The controller monitors the states of tuning tasks and transfers the data between the database side and \sys server side. 
The main parts of \sys run on the server side,  whose resource consumption doesn't affect the online database.
The context featurization module is deployed in the database instance for data privacy concerns.

\noindent\textbf{Discussion.}
\sys is a reactive approach: it featurizes the online workloads at the beginning of an iteration and applies a configuration for the later time of the iteration.
Since changes in real-world workloads are gradual, the little lagging can be acceptable.
A predictive module for the context might help. However, it still assumes gradual and predictable changes based on historical data, and an inaccurate prediction module could greatly affect the tuner's performance.

\section{Top-Level Algorithm}\label{sec:safe-top}
Algorithm \ref{alg:top} presents the top-level algorithm.
\sys first queries the default configuration and its performance to form the initial safety set and initializes the data repository (Line\ref{L1}--\ref{L3}).
The safety threshold is set as the default performance.
At the beginning of an iteration, \sys collects the incoming workload queries and corresponding optimizer's estimations to calculate context vector $c_i$ (Line \ref{alg:context}, \autoref{sec:context-feature}). 
\sys loads a prediction model $m^n_{i-1}$ selected by a SVM model that inputs the context vector and outputs a cluster label $n$ (Line~\ref{alg:svm}, \autoref{sec:context-cluster}). 
Then, a configuration subspace is initialized or adapted (Line \ref{alg:update}, \autoref{sec:safe-update}). 
\sys discretizes the configuration subspace and assesses the safety of unevaluated configurations based on the black-and-white prior knowledge to form a safe candidate set (safety set) (Line \ref{alg:safe}, \autoref{sec:safe-assess}). 
Next, \sys recommends a configuration $\theta_i$ within the safety set to trade-off  exploitation (i.e., making decisions based on
existing knowledge) and exploration (i.e., acquiring new knowledge or expanding the safety set) (Line \ref{alg:select}, \autoref{sec:safe-opt}). 
The configuration $\theta_i$ is applied to the online database, and the database performance $y_i$ during the tuning interval is collected. 
Finally, \sys updates  the prediction model  and data repository  with $\{\theta_i,c_i,y_i\}$ (Lines \ref{alg:model} to \ref{alg:data}).
\sys determines whether to re-cluster or not ((Line \ref{alg:context-spilt}, see Algorithm \ref{alg:context-space} for details).
If needing re-cluster, \sys clusters the observations based on $\{c_i\}_1^{t}$, fits prediction models for each cluster, and re-train the SVM model, as illustrated in \autoref{sec:context-cluster}. 

\begin{algorithm}[h]
\DontPrintSemicolon
\KwOut{Configuration recommendation adaptively with changing environment}
\caption{Top-Level  Algorithm of \sys}\label{alg:top}

 \SetKwFunction{FMain}{Main}
  \SetKwFunction{context}{Offline\_Clustering}
  \SetKwFunction{configuration}{Subspace\_Adaptation}
    \SetKwFunction{append}{Append}
Featurize the environment factor, get context $c_0$.\;\label{L1}
Query default configuration  $\theta_{0}$ and get its performance $y_{0}$.\;
Initialize a data repository $H_0$ with $<c_0, \theta_0, y_0>$. \;\label{L3}
\For{$i \leftarrow$ 1 to $K$}
 {
     Featurize and get context $c_i$.\;\label{alg:context}
     Select a model $m^n_{i-1}$, where $n=SVM(c_i)$.\;\label{alg:svm}
     $\Theta^n_{i}$ = \configuration($\Theta^n_{i-1}, H_{i-1}$).\;\label{alg:update}
     Generate safe candidates $S^n_i \in \Theta^n_{i}$. \;\label{alg:safe}
     Select a configuration $\theta_i$ within $S^n_i$ .\;\label{alg:select}
     Apply  $\theta_i$ and evaluate its performance $y_i$.\;\label{alg:apply}
     $H_i=\append(H_{i-1}, <\theta_i,c_i,y_i>)$. \;\label{alg:data}
     Fit prediction model $m^n_{i}$ on $H_i$.\;\label{alg:model}
      
     \context{$H_i$}.\;\label{alg:context-spilt}
     
 }
\end{algorithm}
\section{Details about subspace adaptation}

\begin{table*}[]
\caption{Average time breakdown for one tuning iteration on JOB workload.}\label{tab:time}
\scalebox{0.9}{
\begin{tabular}{@{}cccccccc@{}}

\toprule
             & Featurization & Model Selection & Model Update & Subspace Adaptation & Safety Assessment & Candidate Selection & Apply \& Evaluation \\ \midrule
Average Time & 0.0577s        & 0.0259s          & 1.3583s       & 0.2695s              & 0.1354s            & 0.0278s              & 178.1254s            \\
Percentage   & 0.03\%        & 0.01\%          & 0.75\%       & 0.15\%              & 0.08\%            & 0.02\%              & 98.96\%             \\ \bottomrule
\end{tabular}}
\end{table*}

 \subsection{More  Rationale about Subspace Restriction}
Given a configuration space $\Theta$ and a contextual GP model, we can discretize $\Theta$ to obtain a set of candidates and assess the safety of each candidate based on the estimation of the GP model.
However, the direct operation over the whole space suffers from the curse of dimensionality. 
First, the candidates need to be close to each other in order to use the GP to generalize safety.
When tuning more than a couple of knobs, it is computationally affordable to conduct a fine discretization over the whole configuration space~\cite{DBLP:conf/icml/KirschnerMHI019}. 
In addition, the database performance function is often complex, making the estimation of the GP model problematic for high dimensional configuration (more than ten parameters)~\cite{DBLP:journals/pieee/ShahriariSWAF16}.
To deal with those challenges, we draw inspiration from a class of trust-region methods in stochastic optimization ~\cite{Yuan_areview}. 
These methods use a linear and quadratic prediction model inside a trust region, which is often a sphere or a polytope centered at the best observation.
Intuitively, while linear and quadratic models are likely to be inadequate to model globally, they can be accurate in a sufficiently small trust region ~\cite{DBLP:conf/nips/ErikssonPGTP19}.
However, the linear models struggle to handle noisy observations and require small trust regions to provide accurate modeling behavior.
Naturally, we use contextual GP as the prediction model to describe the function $f(\theta, c)$ and restrict \sys's optimization space in the trust region (i.e., configuration subspace). 
The optimization can be solved efficiently within the trust region, and the GP model's estimation can be trusted in the trust region ~\cite{DBLP:conf/nips/ErikssonPGTP19}.

\subsection{Direction Oracles for Line Region}
\sys implements two strategies to  generate the directions, including random direction and
important direction.  

\noindent\textbf{Random direction.} A strategy is to pick the direction uniformly (random direction), increasing the exploration.

\noindent\textbf{Important direction.} \sys also chooses the directions aligned with the important configuration knob (important direction), inspired by the important knobs pre-selecting procedure before tuning used by existing configuration tuning approaches ~\cite{DBLP:conf/sigmod/AkenPGZ17, DBLP:conf/kdd/FekryCPRH20,DBLP:conf/hotstorage/KanellisAV20}. 
It is empirically shown that restricting the optimization space in several important knobs can largely reduce the tuning iterations and achieve similar improvement~\cite{DBLP:conf/hotstorage/KanellisAV20}. The importance of knobs is quantified by Fanova~\cite{DBLP:conf/icml/HutterHL14}, a line-time approach for assessing feature importance. However, detecting important knobs needs thousands of evaluation samples for a given workload. 
And fixing the configuration space wrongly (e.g., filtering important knobs) will severely hinder the optimization. 
\sys's adaptation of subspace  solves this problem by adjusting the subspace over iterations.

A random direction is chosen if the performance improvement in the previous hypercube region is lower than a threshold (exploration). 
Otherwise, an important direction is chosen by sampling from the top-5 important knobs (exploitation). 
The importance of knobs is updated with the increasing observations.

\end{document}